\pdfoutput=1
\documentclass[twoside,twocolumn]{article}
\usepackage{blindtext} 

\usepackage[sc]{mathpazo} 
\usepackage[T1]{fontenc} 
\usepackage{microtype} 

\usepackage{cite}
\usepackage{amsmath,amssymb,amsfonts}
\usepackage{algorithmic}
\usepackage{graphicx}
\usepackage{extarrows}
\usepackage{multirow}
\usepackage{threeparttable}
\usepackage{array}
\usepackage{amssymb}
\usepackage{textcomp}
\usepackage{epstopdf}
\usepackage{threeparttable}
\usepackage{subfigure}

\usepackage[english]{babel} 

\usepackage[hmarginratio=1:1,left=18mm,right=18mm,top=32mm,columnsep=20pt]{geometry} 
\usepackage[hang, small,labelfont=bf,up,textfont=it,up]{caption} 
\usepackage{booktabs} 

\usepackage{lettrine} 

\usepackage{enumitem} 
\setlist[itemize]{noitemsep} 

\usepackage{abstract} 

\usepackage{titlesec} 
\renewcommand\thesection{\Roman{section}} 
\renewcommand\thesubsection{\roman{subsection}} 
\titleformat{\section}[block]{\large\scshape\centering}{\thesection.}{1em}{} 
\titleformat{\subsection}[block]{\large}{\thesubsection.}{1em}{} 

\usepackage{fancyhdr} 
\usepackage{titling} 

\usepackage{hyperref} 

\pretitle{\begin{center}\Huge\bfseries} 
\posttitle{\end{center}} 
\title{Deep Attention-guided Hashing} 
\author{%
\textsc{Zhan Yang$^1$, Osolo Ian Raymond$^1$, WuQing Sun$^1$, Jun Long$^{1,2}$}\thanks{Corresponding author. This work was supported in part by the National Natural Science Foundation of China (61472450), the Key Technology R\&D Program of Hunan Province (2018GK2052) and the Science and Technology Plan of Hunan (2016TP1003).} \\[1ex] 
\thanks{\copyright 2018 IEEE. Personal use of this material is permitted. Permission
from IEEE must be obtained for all other uses, in any current or future
media, including reprinting/republishing this material for advertising or
promotional purposes, creating new collective works, for resale or
redistribution to servers or lists, or reuse of any copyrighted
component of this work in other works. See \url{http://www.ieee.org/publications_standards/publications/rights/index.html} for more information}
\thanks{\textbf{IEEE ACCESS} DOI: 10.1109/ACCESS.2019.2891894}
\normalsize $^1$School of Information Science and Engineering, Central South University, Changsha 410083, China \\ 
\normalsize $^2$Network Resources Management and Trust Evaluation Key Laboratory of Hunan Province \\ 
\normalsize \href{mailto:junlong@csu.edu.cn}{junlong@csu.edu.cn} 
}
\date{\today} 
\renewcommand{\maketitlehookd}{%
\begin{abstract}
With the rapid growth of multimedia data (e.g., image, audio and video etc.) on the web, learning-based hashing techniques such as Deep Supervised Hashing (DSH) have proven to be very efficient for large-scale multimedia search. The recent successes seen in Learning-based hashing methods are largely due to the success of deep learning-based hashing methods. However, there are some limitations to previous learning-based hashing methods (e.g., the learned hash codes containing repetitive and highly correlated information). In this paper, we propose a novel learning-based hashing method, named Deep Attention-guided Hashing (\textbf{DAgH}). \textbf{DAgH} is implemented using two stream frameworks. The core idea is to use guided hash codes which are generated by the hashing network of the first stream framework (called first hashing network) to guide the training of the hashing network of the second stream framework (called second hashing network). Specifically, in the first network, it leverages an attention network and hashing network to generate the attention-guided hash codes from the original images. The loss function we propose contains two components: the semantic loss and the attention loss. The attention loss is used to punish the attention network to obtain the salient region from pairs of images; in the second network, these attention-guided hash codes are used to guide the training of the second hashing network (i.e., these codes are treated as supervised labels to train the second network). By doing this, \textbf{DAgH} can make full use of the most critical information contained in images to guide the second hashing network in order to learn efficient hash codes in a true end-to-end fashion. Results from our experiments demonstrate that \textbf{DAgH} can generate high quality hash codes and it outperforms current state-of-the-art methods on three benchmark datasets, CIFAR-10, NUS-WIDE, and ImageNet. 
\end{abstract}
}

\begin{document}
\maketitle


\section{Introduction}
\label{sec:introduction}
In recent years, the amount of multimedia data (text, image, audio and video data) has been growing exponentially. In order to solve the problems of huge storage requirements and learning capacity in dealing with big data, hashing has been the most popular technique for effective binary representation in many tasks due to its fast retrieval and storage efficiency. Generally speaking, the hashing technique, a widely-studied solution to approximate nearest neighbor search, aims to map the original high-dimensional features to a low-dimensional representation, or a short code, called hash code. Then, re-ranking these short codes (hash codes) in response to each query task, requires only a few computations of the Hamming distance operation for efficient multimedia retrieval (i.e., the hashing technique can use a few Bytes to encode one image of several MBytes or one video of several GBytes). Due to the advantages above, hashing has been applied to many large-scale image retrieval~\cite{b1,b2,b3,b4}, text-image cross-model retrieval~\cite{b5}, and person re-identification tasks~\cite{b6}. There are two categories of hashing: data-independent and data-dependent hashing. In this paper, we will build data-dependent hashing methods for generating high quality hash codes, which can capture the potential image representations to achieve better performance than data-independent hashing methods, e.g., Spectral Hashing (SH)~\cite{b7}.

Data dependent methods can be further categorized into supervised and unsupervised methods. Unsupervised methods retrieve the neighbors under some kinds of distance metrics, e.g., Iterative Quantization (ITQ)~\cite{b8}. Compared to the unsupervised methods, supervised methods utilize the semantic labels to improve performance. Many researchers have demonstrated that labels of datasets can improve the quality of hash codes and achieve some success along this direction, e.g., Supervised Hashing with Kernels (KSH)~\cite{b9}, Distortion Minimization Hashing (DMS)~\cite{b4}, Minimal Loss Hashing (MLS)~\cite{b11}, Order Preserving Hashing (OPH)~\cite{b12}, Hamming Distance Metric Learning~\cite{b13}, Semantic Hashing~\cite{b14}, Supervised Discrete Hashing (SDH)~\cite{b15}. However, the quality of hash codes generated is highly dependent on the way feature selection is done, and these methods use hand-crafted features for representation. The need to perform manual feature selection has been a big limitation to the success of these methods.

In the last few years, deep learning networks (e.g., convolutional neural networks) have been shown to have powerful feature extraction capabilities in image processing. They are able to extract high-level features, which leads to attaining much higher performance levels than using hand-crafted features in many image tasks. To solve the limitations of traditional data-dependent hashing methods, this paper focuses on a learning-based hashing method that adopts deep neural networks as the nonlinear functions to enable end-to-end learning of learnable representations and hash codes. These learning-based methods~\cite{b10,b16,b17,b18},  which use pairwise labels to jointly learn similarity-preserving representations and optimize the pair-wise loss and quantization loss, have exhibited high performance on many benchmark tests.

\begin{figure}
    \centering
    \includegraphics[width=8cm]{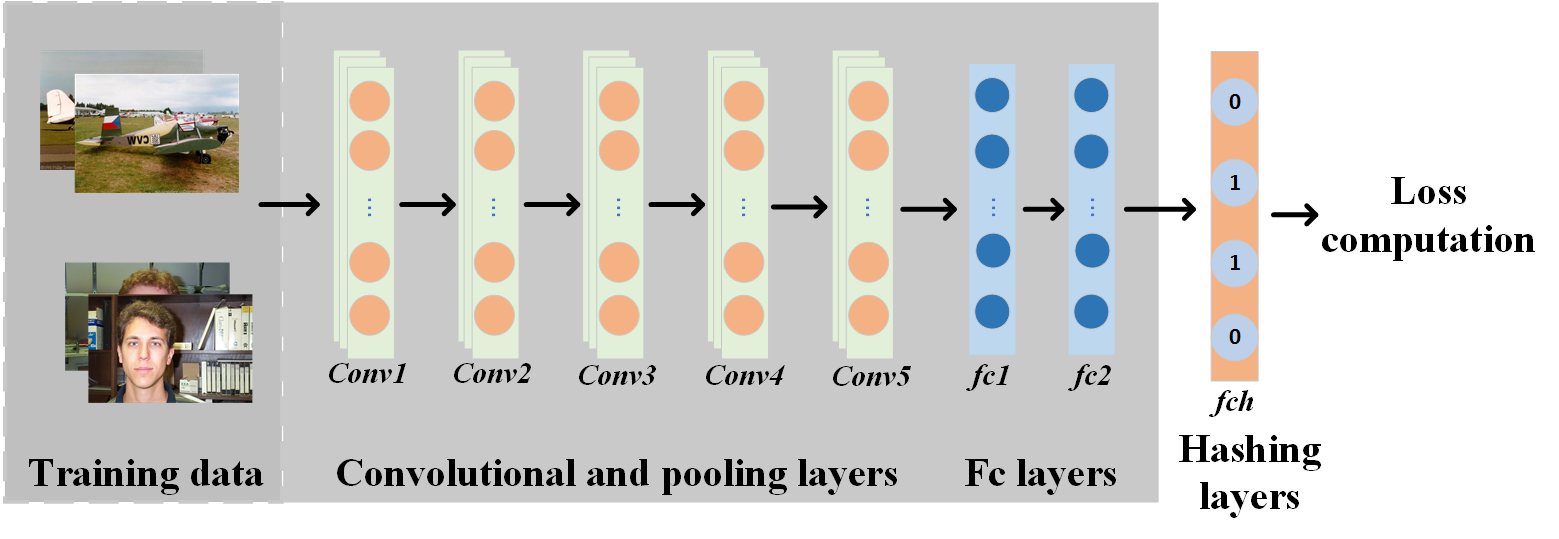}\\
    \caption{The basic architecture of supervised learning-based hashing.}\label{fig:F1}
  \end{figure}

Although recent learning-based hashing methods have achieved significant progress in multimedia retrieval, there are some limitations of previous learning-based hashing methods in generating long hash codes (say, more than 24 bits), e.g., the learned long hash codes contain repetitive and highly correlated information. Any natural image will contain some useless information, or some interference information that is not relevant for a particular task. Images of the same category may contain completely different backgrounds, different categories of images may have similar backgrounds, directly generating the hash codes by a standard learning-based method (as shown in Figure~\ref{fig:F1}) will in practice result in a higher possibility of having correlated bits as the length of the hash codes increases. Then, highly correlated bits have a large impact on retrieval performance (i.e., the cost-performance ratio decreases with increase in the length of the hash code). As an extreme example, if 256-bit hash codes are positively and negatively completely correlated, the performance will be similar to that of the 1-bit hash codes. To solve these limitations, in this paper, we deal with the salient regions and backgrounds of the images separately. Specifically, the main idea of the paper is to firstly adopt an attention network to generate the attention images from the original images, (i.e, use visual attention models to localize regions in an image to capture features of the regions) and then use pairwise information to generate the attention-guided hash codes from the attention images. Secondly, we use these attention-guided hash codes to guide the training of the second hashing network (i.e., these codes are treated as supervised labels to train the second network).

The contributions of this work are summarized as follows:
\begin{enumerate}
\item The proposed \textbf{DAgH} model combines two stream frameworks. The first stream framework consists of an attention network and a hashing network (the first hashing network). The role of the first stream framework is to generate the attention-guided hash codes. A novel method of using the semantic loss and attention loss to train the first stream framework is proposed. The second stream framework contains another hashing network i.e. the second hashing network. This hashing network is guided by the attention-guided hash codes which were generated from the first stream framework. The second stream framework is trained by the proposed guide loss. To the best of our knowledge, this is the first learning-based hashing method that uses its own attention-guided hash codes to guide the training of the original image hashing network.

\item In order to guarantee the quality of the final hash codes and eliminate the quantization error, the \textbf{DAgH} model uses a continuous activation function to ensure that the first stream framework is a true end-to-end network and a threshold activation function to ensure that the second stream framework directly generates the final hash codes. In the first stream framework, we chose to use a continuous ATanh activation function for training because it's easier to optimize than using a $sign$ function with no extra quantization loss. As a result, it shows stronger capacity in learning high quality attention-guided hash codes. When the second stream framework is trained by the attention-guided hash codes, we can use a $sign$ activation function to constrain the output of the second stream framework for generating the binary codes directly. These operations trade off efficacy for efficiency.
\end{enumerate}

The remainder of this paper is structured as follows. In Section \ref{sec:Related_Work}, we briefly introduce the related works. In Section \ref{sec:Method}, we highlight the motivation of our method and provide some theoretical analysis for its implementation. In Section \ref{sec:EXPERIMENTS}, we introduce our experimental results and corresponding analysis and finally in Section \ref{sec:Conclusion_and_Future_Work} conclude the paper.

\section{Related Work}\label{sec:Related_Work}
\subsection{Hashing}
By representing multimedia data as binary codes and taking advantage of fast query retrieval, hashing is a novel technique that can resolve the information retrieval problems in this multimedia era. Wang \emph{et al.}~\cite{b21} have provided a comprehensive literature survey that covers the most important methods and latest advances in image retrieval.

We can divide the hashing methods into two categories: data-independent and data-dependent methods. In the early researches, Locality Sensitive Hashing (LSH) was one of the data-independent methods used. LSH hashes input items so that items that are similar have a high probability of being mapped to the same ``buckets'' (the number of buckets being much smaller than the universe of possible input items)~\cite{b22}. LSH and several variants (e.g., kernel LSH~\cite{b23} and $p$-norm LSH~\cite{b24}) are widely used for large-scale image retrieval. However, there are many limitations of data-independent methods, e.g., the efficiency is low and it requires longer hash codes to attain high performance. Due to the limitations of the data independent methods, current researchers focus on using a variety of machine learning techniques to learn more efficient hash functions based on a given dataset.

Data dependent methods can be further categorized into supervised, semi-supervised and unsupervised methods. Unsupervised hashing methods learn hash functions that encode data points to binary codes by training from unlabeled data. Typical learning criteria include minimize reconstruction error~\cite{b25,b26,b27,b28} and graph structure learning~\cite{b29,b30}. Iterative Quantization (ITQ) is one of the unsupervised methods in which the projection matrix is optimized by iterative projection and thresholds according to the given datasets~\cite{b8}. Compared to the semi-supervised and unsupervised methods, supervised methods utilize the semantic labels to improve performance. Many researchers have proposed along this direction and have achieved some success (They have demonstrated that labels of datasets can improve the quality of hash codes), e.g., Supervised Hashing with Kernels (KSH)~\cite{b9}, Distortion Minimization Hashing (DMS)~\cite{b4}, Minimal Loss Hashing (MLS)~\cite{b11}, Order Preserving Hashing (OPH)~\cite{b12}, Hamming Distance Metric Learning~\cite{b13}, Semantic Hashing~\cite{b14}, Supervised Discrete Hashing (SDH)~\cite{b15}. The hash codes are generated by minimizing the Hamming distance between similar pairs and maximizing the Hamming distance between dissimilar pairs.

Recently, deep convolutional neural networks have yielded amazing results on many computer vision tasks, this success has attracted the attention of researches of learning-based hashing methods. Convolutional Neural Network Hashing (CNNH) is one of the early works to use a learning-based hashing method, which utilize two stages to learn the image features and hash codes. Following this work, many learning-based hashing techniques have been proposed, e.g., Deep Semantic Ranking Hashing (DSRH)~\cite{b31} which learns the hash functions by preserving semantic similarity between multi-label images. Deep Visual-Semantic Quantization (DVSQ)~\cite{b1} generates the compact hash codes by optimizing an adaptive margin loss and a visual-semantic quantization loss over multi-networks. Deep Supervised Hashing (DSH)~\cite{b32} designs a loss function to pull the outputs of similar pairs of images together and pushes the dissimilar ones far away. Its outputs are relaxed to real values to avoid optimizing the non-differentiable loss function in Hamming distance. Network In Network Hashing (NINH)~\cite{b33} adopts a triplet ranking loss to capture the relative similarities of images. Deep Supervised Discrete Hashing (DSDH)~\cite{b34} uses both pairwise label and classification information to learn the hash codes under a single steam framework. Guo \emph{et al.,}~\cite{b35} show that existing DSH can achieve good results with short hash codes (e.g., 8 to 24 bits) but only lead to marginal performance gain with long hash codes (e.g., 128 bits). They try to divide a single network into many sub-networks to generate hash codes respectively. Extensive researches have taken advantage of deep learning techniques to achieve great improvements compared to traditional data-dependent hashing methods.

However, existing learning-based methods do not consider the high correlation problem of long hash codes. Although convolutional neural networks have powerful capabilities in image feature extraction, they do not deal with the irrelevant features in the image. When long hash codes need to be generated, the correlation problem of the hash codes cannot be ignored. In this paper, we introduce a high-quality hash code generation method, where an attention network is embedded to mine salient regions for guiding the standard supervised learning-based hashing framework.

\subsection{Salient Regions Learning}
The key challenge of learning high quality hash codes is to locate the salient regions in images. Many methods for locating salient regions have been proposed in recent years. Previous methods of locating the salient regions can be categorized into traditional methods and deep learning based methods.

Traditional methods include techniques such as~\cite{b36,b37,b38} locating the salient regions by unsupervised methods. Following these works, some hashing methods locate salient regions to improve performance in the unsupervised manner. Shen \emph{et al.}~\cite{b39} proposed a cross-modal hashing method which uses RPN~\cite{b40} to detect salient regions. Then, the two cross-modal networks are used to encode the region information, the semantic dependencies and cues between the words. DPH~\cite{b41} uses GBVS~\cite{b42} to count the scores for each pixel. Then, a collection of salient regions are generated based on increasing threshold scores. However, traditional methods use ready-made models to locate salient regions and therefore, when encountering a new dataset, there is no guarantee that the learned salient regions are accurate.

Due to the success of deep learning, most of the methods depend on powerful deep features, which have shown a higher performance gain than hand-crafted features on image classification~\cite{b43,b44,b45,b46}. Zhao \emph{et al.}~\cite{b47} adopted similarity labels to train part model for person re-identification. Lin \emph{et al.}~\cite{b48} proposed a bilinear structure, which computes the pairwise feature interactions by two-stream convolutional neural networks to capture the different salient regions between input images and achieved high performance in bird classification. RA-CNN~\cite{b49} is a recurrent-attention convolutional neural network, which can discover salient regions and learn region-based features recursively. Motivated by~\cite{b49,b50}, we adopt a novel attention network to generate the attention image and use a hashing network to learn the attention-guided hash codes. Then use the generated attention-guided hash codes to guide the second hashing network to learn the final hash codes.

\begin{figure*}
    \centering
    \includegraphics[scale=0.305]{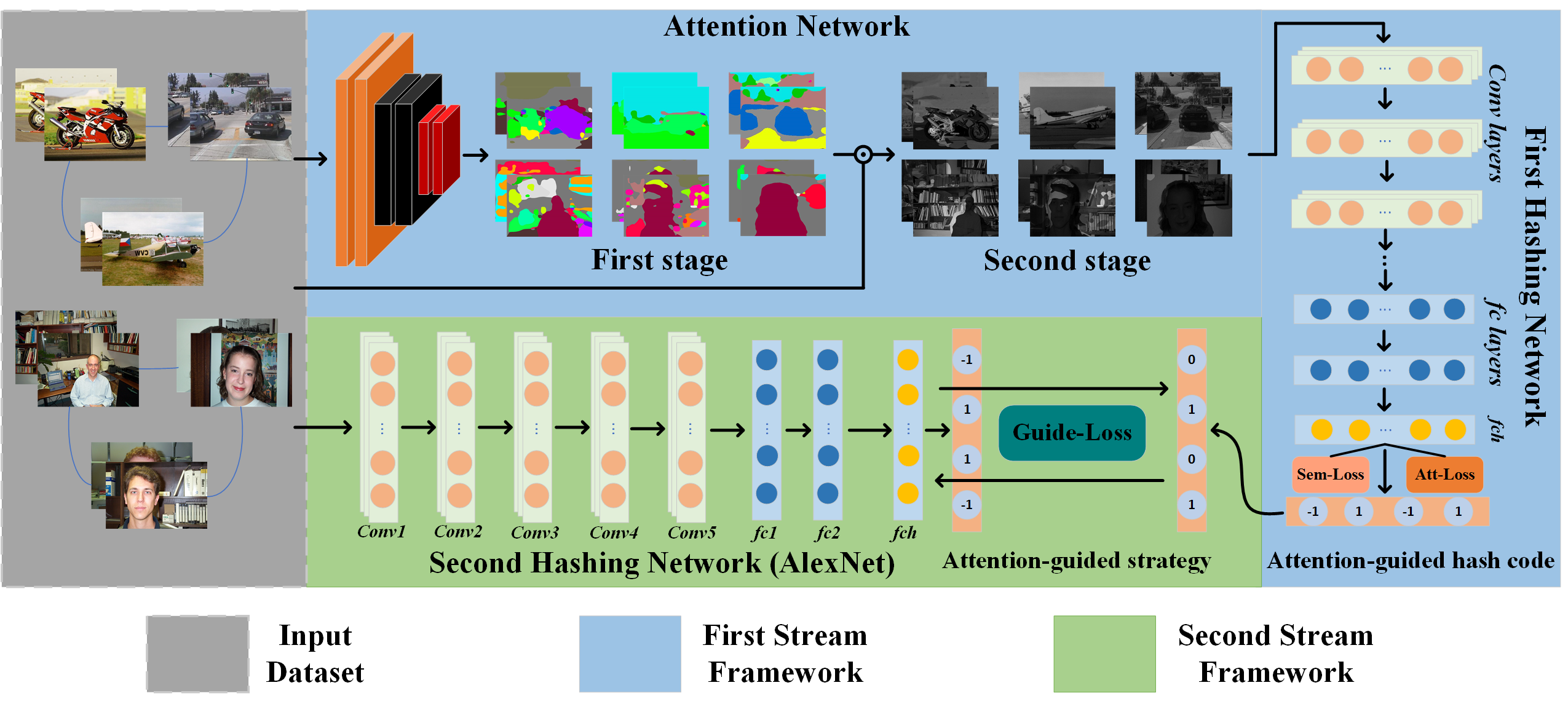}\\
    \caption{The proposed architecture for deep attention-guided hashing (\textbf{DAgH}). \textbf{DAgH} consists of two stream frameworks: 1) the first stream framework contains an attention network based on FCN-16 network for learning the attention image pair, the attention processing contains two stages. Then, the hashing network uses AlexNet (or ResNet) for learning the attention-guided hash codes. The first stream framework consists of two loss functions: the semantic loss and the attention loss, and uses a continuous ATanh activation function for training. 2) the second stream framework contains a hashing network that adopts AlexNet for learning hash codes, it then uses the attention-guided hash codes for its supervised labels, and the final hash codes are generated directly by the second stream framework using $sign$ activation function. (Best viewed in color.)}\label{fig:F2}
  \end{figure*}

\section{Deep Attention-guided Hashing}\label{sec:Method}
In this section, we first give the problem formulation, then show the details of our proposed method, including the framework, loss function and training strategy, and finally show its extensions to out-of-sample data.

\subsection{Problem Formulation}

In similarity retrieval systems, we are given a training set $\mathcal{X}=\{x_i\}_{i=1}^N$, each image represented by a $d$-dimensional feature vector $x_i\in\mathbb{R}^d$, where $\mathcal{X}\in\mathbb{R}^{d\times N}$. In supervised learning-based hashing, the pairwise information $S=\{s_{ij}\}$\footnote{Note that one image may belong to multiple categories.} is derived as:
\begin{equation}\label{eq:pairwise_label}
  s_{ij}=\begin{cases}
           1, & \mbox{if } \text{images}\ x_i\ \text{and}\ x_j\ \text{share\ same\ class label}\\
           0. & \mbox{otherwise}
         \end{cases}
\end{equation}

Supervised learning-based hashing method learn a nonlinear function $f:\ \mapsto b\in\{-1,1\}^K$ from an input space $\mathbb{R}^d$ to Hamming space $\{-1,1\}^K$ with deep neural network. This method generally contains three steps: 1) using a network for learning deep features of each image $x_i$, 2) using a fully-connected hashing layer ($fch$) for transforming the deep features into $K$-dimensional continuous representation $\omega_i\in\mathbb{R}^K$, 3) using a $sign$ function to quantize the continuous representation $\omega_i$ into $K$-bit binary hash code $b_i\in\{-1,+1\}^K$. The similarity labels $S=\{s_{ij}\}$ can be constructed from semantic labels of data points or relevance feedback in real retrieval systems. In addition, the threshold function $sign(\cdot)$ is an element-wise $sign$ function defined as follows:
\begin{equation}\label{eq:sign_f}
  sign(x)=\begin{cases}
            1, & \mbox{if}\ \ x\geq 0\\
            -1. & \mbox{otherwise}
          \end{cases}
\end{equation}

\subsection{Network Architecture}
To address the limitations of previous learning-based hashing methods, we propose a novel learning-based method. Figure~\ref{fig:F2} shows the proposed \textbf{DAgH} model. Our method includes two stream framework. The first stream contains an attention network and a hashing network. The attention map is the most critical part of our network, since it allows the network to know which regions should be focused on. Then the hashing network uses the attention images to generate the attention-guided hash codes. Thereafter, in the second stream, these attention-guided hash codes are treated as supervised labels to train an image hashing network which can generate the final hash codes for the input images. The details of all stream frameworks are described in the following subsections.

\subsection{The First Stream Framework}\label{sec:The_First_Attention-Hashing_Network}
As mentioned previously, this stream framework is a true end-to-end deep model which includes an attention network, i.e., $Attention(\cdot|\Theta_{a})$, and a hashing network, i.e., $Hash(\cdot|\Theta_1)$. In this subsection, we will introduce the main processes of the two networks and the details of the network training.

\textbf{The Attention Network} The role of the attention network is to find salient regions in the original image that need to get attention. These regions are the most representative of the theme of the image, so they can be used for better salient region restoration, and for the first hashing network to focus the assessment on. This attention processing consists of two stages. In the first stage, the proposed FCN-based attention network~\cite{b61} is used to map the original input image pair $[x_i,x_j]$ to the preliminary attention image pair $[\hat{x}_i,\hat{x}_j]$. Inspired by~\cite{b50}, to build a learnable attention network and guarantee that it generates accurate attention images, we define a normalization function to restrict the value of each pixel between 0 and 1:
\begin{equation}\label{eq:normalize_pixel}
  norm_i(p,q)=\frac{x_i(p,q)-\text{min}(x_i)}{\text{max}(x_i)-\text{min}(x_i)},
\end{equation}
where $(p,q)$ denotes the location $(p,q)$ in an image $x$, $\text{max}(x)$ denotes the maximum pixel value in an image $x$, $\text{min}(x)$ denotes the minimum pixel value in an image $x$.

In the second stage, the attention image pair $[\tilde{x}_i,\tilde{x}_j]$ is computed through a Hadamard product $\otimes$ of the original image pair $[x_i,x_j]$ and the normalization function of the preliminary attention image pair $[\hat{x}_i,\hat{x}_j]$:
\begin{equation}\label{eq:finally_attetion_image}
  [\tilde{x}_i,\tilde{x}_j]=[x_i,x_j]\otimes norm_{[\hat{x}_i,\hat{x}_j]}(p,q).
\end{equation}

Then we can encode the attention image pair $[\tilde{x}_i,\tilde{x}_j]$ by the first hashing network. The attention network can be gradually fine-tuned their parameters through~\eqref{eq:overall_loss} to mine salient regions automatically.

\textbf{The First Hashing Network} The generated attention image pair $[\tilde{x}_i,\tilde{x}_j]$ as the input of the first hashing network and the semantic information as the pairwise label to train the first hashing network. After the first hashing network is trained, the attention-guided hash code  $B^{att}=\{b_i^{att}\}^N_{i=1}$ is calculated through the trained hashing network:
\begin{equation}\label{eq:first_hashing}
  b_i^{att}=sign(Hash(\tilde{x}_i|\Theta_1)),
\end{equation}
where $b_i^{att}$ is the attention-guided hash code, $\Theta_1$ denotes the parameters of the first hashing network\footnote{Note that, the ATanh activation function(the detail about the ATanh activation function can be found in~\eqref{eq:Atanh}) is used to train the first stream framework~\eqref{eq:loss_function_x} and the $sign$ activation function is used as the final output of the first stream framework.}, and $\tilde{x}_i$$=$$Attention(x_i|\Theta_{a})$ is the attention image that is input into the first hashing network and generated by the attention network, $\Theta_{a}$ is the parameters of the attention network.

The no-quantization loss training strategy of the first stream framework is expatiated as follows:

\noindent\textbf{Similarity Measure}

\noindent For a pair of binary hash codes $b_i$ and $b_j$, the relationship between their Hamming distance $\text{dist}_H$ and inner product $\langle\cdot,\cdot\rangle$ is formulated as follows: $\text{dist}_H=\frac{1}{2}(K-\langle b_i,b_j\rangle)$. The larger the inner product of two hash codes, the smaller the Hamming distance, and vice versa. Therefore, the inner product through two hash codes is a reliable criterion for evaluating the similarity between them.

In supervised learning-based hashing method, the Maximum Likelihood (WL) estimation of the hash codes $B=[b_1,b_2,...,b_N]$ for all $N$ images is:
\begin{equation}\label{eq:WL}
  \text{log}P(S|B)=\prod_{s_{ij}\in\mathcal{S}}\text{log}P(s_{ij}|B),
\end{equation}
where $P(S|B)$ denotes the likelihood function. Given each image pair with their similarity label $([x_i,x_j],s_{ij})$, $P(s_{ij}|b_i,b_j)$ is the conditional probability of $s_{ij}$ given the pair of corresponding hash codes $[b_i,b_j]$, which is naturally defined as logistic function:
\begin{equation}\label{eq:logistic_function}
  P(s_{ij}|b_i,b_j)=\begin{cases}
                      \sigma(\langle b_i,b_j\rangle), &  s_{ij}=1\\
                      1-\sigma(\langle b_i,b_j\rangle), &  s_{ij}=0
                    \end{cases}
\end{equation}
where $\sigma(x)=1/(1+e^{-x})$ is the sigmoid activation function, $\langle b_i,b_j\rangle=\frac{1}{2}b_i^Tb_j$.

\noindent\textbf{Loss Function}

\noindent \textbf{Semantic Loss} Considering the similarity measure,the following loss function is used to learn the hash codes:
\begin{equation}\label{eq:loss_function_x}
\begin{aligned}
  \mathcal{L}_{sem}&=-\log P(S|B)=-\sum_{s_{ij}\in S}\log (s_{ij}|B)\\
  &=-\sum_{s_{ij}\in S}(s_{ij}\langle b_i,b_j \rangle-\log(1+\text{exp}(\langle b_i,b_j \rangle))),\\
\end{aligned}
\end{equation}
where $b_i=sign(\omega_i)$, which converts the $K$-dimensional representation $\omega_i$ to exactly binary hash codes\footnote{Note that, Equation~\eqref{eq:loss_function_x} need to first learn the continuous representation $\omega_i$, which are quantized to binary values in a separated operation using $sign$ function, this will result in quantization errors.}. Equation~\eqref{eq:loss_function_x} is the negative $log$ likelihood loss function, which represents the Hamming distance of two similar images that are as small as possible, and the Hamming distance of two dissimilar images that are as large as possible. Then, we define an attention loss to train the attention network to capture some salient regions of the image.

\textbf{Attention Loss} In training the attention network, we denote the continuous representation pair of $fch$ layer (also called binary-like codes) as $[\omega_i,\omega_j]$. Then we obtain the optimal hash code pair $[b_i,b_j]$ from the continuous representation pair $[\omega_i,\omega_j]$. Given $[b_i,b_j]\in\{-1,1\}^k$, the cosine similarity between the continuous representation pair can be defined as $\cos(\omega_i,\omega_j)=\frac{\omega_i^T\omega_j}{||\omega_i||_2||\omega_j||_2}$, which is in the range of $(-1,1)$. Therefore, we use $\frac{\cos(\omega_i,\omega_j)+1}{2}$ to restrict the similarity value to $(0,1)$. The attention loss is written as below:
\begin{equation}\label{eq:attention_loss}
\begin{aligned}
  \mathcal{L}_{att}&=\sum_{i,j}||S_{ij}-\frac{\cos(\omega_i,\omega_j)+1}{2}||_2\\
  &+\sum_{i,j}\max(0,\lambda-||S_{ij}-\frac{\cos(\omega_i,\omega_j)+1}{2}||_2),\\
\end{aligned}
\end{equation}
where $\lambda>0$ is a margin parameter. The attention loss will punish the attention network to make it better capture the salient regions.

Overall, combining Equation~\eqref{eq:loss_function_x} and~\eqref{eq:attention_loss}, the loss of the first stream framework can be written as:
\begin{equation}\label{eq:overall_loss}
  \underset{\Theta_a,\Theta_1}{\text{min}}\ \mathcal{L}_{sem}+\nu\mathcal{L}_{att},
\end{equation}
where $\Theta_a,\Theta_1$ are the first stream framework parameters efficiently optimized using standard back-propagation with automatic differentiation techniques.

\noindent\textbf{End-to-End Learning}

\noindent In many of the recent hashing methods~\cite{b8,b15,b16,b32,b34,b50,b51,b52,b53}, quantization error is an important part of their optimization process, which will directly result in retrieval quality. These hashing methods first need to learn continuous representations (binary-like codes) through $sigmoid$ and $\tanh$ functions, then, the binary-like codes are binarized into hash codes in a separate operation of $sign$ thresholding. Therefore, the gap between the binary-like codes and hash codes is called the quantization error. For examples, in ~\cite{b50}, the quantization error is defined as $\mathcal{L}_{reg}=\sum_{i}||\omega_i-b_i||_1$, in ITQ~\cite{b8}, the quantization error is defined as $\mathcal{L}_{ITQ}=||\omega_i-b_i||_2$, where $b_i=sign(\omega_i)\in\{1,-1\}^K$. Although the optimization methods propose to reduce the quantization error, the activations of the $fch$ layer are still not binary. This is because a $sign$ function is non-smooth and non-convex, and therefore has no gradient (i.e., the gradient of $sign$ function is zero for all non-zero inputs, which makes the classical back-propagation infeasible for training deep networks.). Cao \emph{et al.}~\cite{b54} proposed a justifiable approach based on the continuation of the $\tanh$ function, which approaches the $sign$ function with the scale parameter $\beta$ in its limit: $\lim_{\beta \rightarrow \infty}\tanh(\beta x)=sign(x)$, they prove the convergence of this optimization when adopting a sequence of increasing values of $\beta$ during training. However, in order to ensure that the continuous $\tanh$ function is differential everywhere that can be optimized via standard back-propagation, a regularization term should be considered~\cite{b55}. Such activation function is named as Adaptive Tanh (ATanh):
\begin{equation}\label{eq:Atanh}
  b_i=\tanh(\beta \omega_i) + \epsilon\ ||\frac{1}{\beta}||_2^2,
\end{equation}
where $\epsilon$ is the regularization constant. The second term of~\eqref{eq:Atanh} is a regularization term. The regularization term is a penalty to the standard $\tanh(\beta x_i)$, when $\beta$ gradually increases, the ATanh function approaches the $sign$ function and has the reliable-ability to generate hash codes. When $\beta\rightarrow\infty$, the optimization problem will converge to the original deep learning to hash problem in~\eqref{eq:loss_function_x} with $sign(x)$ activation function. We follow the empirical parameters setting and first set the parameter $\beta_0=1$ as the initialization. At each epoch $T$, we increase $\beta$ and fine-tune the first stream framework of \textbf{DAgH} in the next epoch. With the parameter $\beta\rightarrow\infty$ of the~\eqref{eq:Atanh}, the network will converge to the first stream framework of \textbf{DAgH} with $sign$ as activation function, which can generate high-quality attention-guided hash codes as we required. The time consumption of ATanh as the activation function in the whole network is negligible (i.e., both forward and backward computation is negligible)~\cite{b55}. Different from the previous hashing methods mentioned above, there is no-extra quantization error within such an end-to-end hashing net, hence it shows stronger capacity in learning high-quality attention-guided hash codes.

\begin{table}
\label{Al:Algorithm 1}
\setlength{\tabcolsep}{3pt}
\begin{tabular}{p{240pt}}
\hline
\specialrule{0em}{2pt}{2pt}
\textbf{Algorithm 1} Deep Attention-guided Hashing (\textbf{DAgH}).\\
\specialrule{0em}{2pt}{2pt}
\hline
\specialrule{0em}{2pt}{2pt}
\textbf{Input} Training Image pair with their similarity label $([x_i,x_j],s_{ij})$ in the first stream framework, a sequence $1=\beta_0<\beta_1<\beta_2...<\beta_m=\infty$. Training Image $x_i$ and the attention-guided hash codes $B^{att}$ in the second stream framework. Training epoches $T_1$ and $T_2$ of the first and second stream framework optimizations, respectively.\\
\specialrule{0em}{2pt}{2pt}
\textbf{Output} First stream framework: $sign(Hash(Attention(x_i|\Theta_{a})|\Theta_1))$; Second stream framework: $sign(Hash(x_i|\Theta_2))$.\\
\specialrule{0em}{2pt}{2pt}
\textbf{Begin} Construct the pairwise information matrix $S$ according to~\eqref{eq:pairwise_label}.\\
\specialrule{0em}{2pt}{2pt}
1.\ \textbf{for} $t=1:T_1$ epoch \textbf{do}\\
\specialrule{0em}{1pt}{1pt}
2.\ \ \ Compute $[b_i^{att},b_j^{att}]$ according to~\eqref{eq:first_hashing}\\
\specialrule{0em}{1pt}{1pt}
3.\ \ \ Train the first hashing network~\eqref{eq:loss_function_x} with~\eqref{eq:Atanh} as activation\\
\specialrule{0em}{1pt}{1pt}
4.\ \ \ Compute $\Theta_{a},\Theta_{1}$ according to~\eqref{eq:overall_loss}\\
\specialrule{0em}{1pt}{1pt}
5.\ \ \ Set converged the first stream framework as next epoch initialization\\
\specialrule{0em}{1pt}{1pt}
6.\ \textbf{end for}\\
\specialrule{0em}{1pt}{1pt}
7.\ \textbf{return} $sign(Hash(Attention(x_i|\Theta_{a})|\Theta_1))$, $\beta_m\rightarrow\infty$.\\
\specialrule{0em}{1pt}{1pt}
\hline
\specialrule{0em}{1pt}{1pt}
1.\ \textbf{for} $t=1:T_2$ epoch \textbf{do}\\
\specialrule{0em}{1pt}{1pt}
2.\ \ \ Compute $\tilde{y}_i^2$ according to $\tilde{y}_i^2=Hash(x_i|\Theta_2)$\\
\specialrule{0em}{1pt}{1pt}
3.\ \ \ Compute $\Theta_2$ according to~\eqref{eq:second_loss_function}\\
\specialrule{0em}{1pt}{1pt}
4.\ \ \ Set converged the second stream framework as next epoch initialization\\
\specialrule{0em}{1pt}{1pt}
5.\ \textbf{end for}\\
\specialrule{0em}{1pt}{1pt}
6.\ \textbf{return} $sign(Hash(x_i|\Theta_2))$.\\
\specialrule{0em}{2pt}{2pt}
\hline
\end{tabular}
\end{table}

\subsection{The Second Stream Framework}
As shown in Figure~\ref{fig:F2}, we directly adopt a pre-trained AlexNet as the base of the second hashing network. After obtaining the attention-guided hash code $B^{att}$, we thereafter utilize it as the supervised labels and the original images $\mathcal{X}=\{x_i\}^N_{i=1}$ to train the hashing network. When the hashing network is trained, the final hash codes $b_i^f$ are computed through the trained hashing network:
\begin{equation}\label{eq:second_hashing}
  b_i^f=sign(Hash(x_i|\Theta_2)),
\end{equation}
where $b_i^f$ is the final hash code, $\Theta_2$ denotes the parameters of the second hashing network, and $\tanh(\cdot)$ as activation of the $fch$ layer of the hashing network.

The details of learning strategy are explained as follows:

\noindent\textbf{Attention-Guided Strategy}

\noindent As mentioned above, when the first stream framework is trained, the generated attention-guided hash codes are used as the supervised labels to guide the second hashing network. Considering the powerful image feature extraction ability of convolutional neural networks, here we adopt the famous and widely used AlexNet, which is commonly used in baseline models. The AlexNet consists of 5 convolutional layers $(c1-c5)$, and 2 fully-connected layers $(fc1-fc2)$, and is pre-trained on the ImageNet dataset. To obtain the hash codes, we add a $k$-nodes hash layer, called $fch$, each node of $fch$ layer corresponds to 1 bit in the target hash code. With the $fch$ layer, the previous layer representation is transformed to a $k$-dimensional representation. The architecture of the second hashing network is shown in Figure~\ref{fig:F2}.

More specifically, let $\tilde{y}_i^2=Hash(x_i|\Theta_2)$ be the output of the second hashing network, where $x_i$ is the original input image and $\Theta_2$ is the parameter of the second hashing network. Since our goal is to use the attention-guided hash code $b_i^{att}$ to guide the second hashing network through sigmoid cross-entropy loss function, we need to convert the value of -1 in the attention-guided hash codes to 0 so that the value of the attention-guided hash code is $b_i^{att}\in\{0,1\}^K$. We define the following likelihood functions:
\begin{equation}\label{eq:AG_1}
  P(b_{ik}^{att}|\tilde{y}_{ik}^2)=\begin{cases}
                                 \sigma(\tilde{y}_{ik}^2), & \ \ \ \ b_{ik}^{att}=1 \\
                                 1-\sigma(\tilde{y}_{ik}^2), & \ \ \ \ b_{ik}^{att}=0
                               \end{cases}
\end{equation}
where $b_{ik}^{att}$ is the hash code corresponding to the $k$-th bit of the $i$-th element in $b_i^{att}$, $\ \tilde{y}_{ik}^2$ is the output of the $k$-th node in $fch$ layer of the $i$-th element, and $\sigma(\cdot)$ is a sigmoid function as shown in~\eqref{eq:logistic_function}.

\noindent\textbf{Loss Function}

\noindent \textbf{Guide Loss} In order to use the attention-guided hash codes to guide the second hashing network, we define a guide loss, which is written as follows:
\begin{equation}\label{eq:second_loss_function}
\begin{aligned}
	&\mathcal{L}_g=-\frac{1}{KN}{\sum_{k=1}^K}\sum_{i=1}^{N}{\log}P\left( b_{ik}^{att}|\tilde{y}_{ik}^{2} \right)\\
    &=-\frac{1}{KN}\sum_{k=1}^K\sum_{i=1}^{N}[{\log P_{ik}}^{b^{att}_{ik}}\cdot{\log(1-P_{ik})}^{(1-b_{ik}^{att})}]\\
	&=-\frac{1}{KN}\sum_{k=1}^K\sum_{i=1}^{N}{\left[ b_{ik}^{att}\log P_{ik}+\left( 1-b_{ik}^{att} \right) \log \left( 1-P_{ik} \right) \right]},\\
\end{aligned}
\end{equation}
where $N$ is the number of training images, $K$ is the number of bits in each hash code, and $P_{ik}=\sigma(\tilde{y}_{ik}^{2})$.

In order to minimize~\eqref{eq:second_loss_function}, we use the Back-Propagation (BP) algorithm to learn the parameter $\Theta_2$ of the second hashing network with stochastic gradient descent (SGD). Specifically, we take the derivative of the guide loss:
\begin{equation}\label{eq:derivation}
\begin{aligned}
  &\frac{\partial \mathcal{L}_g}{\partial \tilde{y}_{ik}^2}=\frac{\partial \mathcal{L}_g}{\partial P_{ik}}\frac{\partial P_{ik}}{\partial \tilde{y}_{ik}^2}\\
  &=-\frac{1}{KN}(b_{ik}^{att}\frac{1}{P_{ik}}-\frac{1-b_{ik}^{att}}{1-P_{ik}})(P_{ik}(1-P_{ik}))\\
  &=-\frac{1}{KN}(P_{ik}-b_{ik}^{att}).
\end{aligned}
\end{equation}

Thereafter, we can obtain $\partial \mathcal{L}_g/\partial \Theta_2$ with $\partial \mathcal{L}_g/\partial \tilde{y}_{ik}^2$ using the chain rule, i.e., we can use BP to update the parameter $\Theta_2$ of the second hashing network. After training, we obtain the trained AlexNet model for the final hashing model and the corresponding image hash codes can be generated by~\eqref{eq:second_hashing}.

\subsection{Out-of-Sample Extension}
After our proposed \textbf{DAgH} model is trained, we can easily generate its hash code through the second hashing network. For example, given a new instance $x_q\notin\mathcal{X}$, we directly use it as the input of \textbf{DAgH} model, then forward propagate the second hashing network to generate its hash code as follows:
\begin{equation}\label{eq:out_of_sample_hashing}
  b_q=sign(Hash(x_q|\Theta_2)).
\end{equation}

\section{Experiments}\label{sec:EXPERIMENTS}
In order to demonstrate the performance of our proposed \textbf{DAgH} method, we carried out extensive experiments on three widely used benchmark datasets, i.e., CIFAR-10, NUS-WIDE, and ImageNet, to verify the effectiveness of our method.

\subsection{Datasets and Settings}
\textbf{CIFAR-10}~\cite{b56} dataset consists of 60,000 images with a resolution of 32$\times$32 in 10 categories (each category has 6,000 images). Each image has only one category. In our experiment, we randomly selected 100 images per category (i.e., 1,000 images in total) as the test set, 500 images per category (i.e., 5,000 images in total) as the training set. The rest of the images are used as gallery in the testing phase.

\noindent\textbf{NUS-WIDE}~\cite{b57} is a dataset contains that nearly 270K (260,648) images collected from the public web. It is a \emph{multi-label} dataset. There are 81 semantic concepts manually annotated for evaluating retrieval performance. In our experiment, as in~\cite{b17} and~\cite{b34}, we selected the 21 most frequent concepts. We randomly sample 100 images per class (i.e., 2,100 images in total) as the test set, 500 images per class (i.e., 10,500 images in total) as the training set. The rest of the images are treated as the gallery in the testing phase.

\noindent\textbf{ImageNet}~\cite{b58} dataset is a well-known benchmark dataset for the Large Scale Visual Recognition Challenge (ILSVRC 2015). It contains 1,000 categories with over 1.2M images in the training set and 50,000 images in the validation set, where each image has only one category. As in~\cite{b3} and~\cite{b53}, we randomly selected 100 categories which led to a database with about 120K images and a query set with about 5,000 images. In this dataset, we randomly selected 100 images per class (i.e., 10,000 in total) as the training set.

\subsection{Baselines}
We compared our proposed \textbf{DAgH} method against some classic or state-of-the-art hashing methods. We roughly divided these methods into two groups: traditional hashing methods and learning-based hashing methods. The traditional hashing methods include unsupervised hashing methods: \textbf{SH}~\cite{b7}, \textbf{ITQ}~\cite{b8}, and supervised hashing methods: \textbf{SDH}~\cite{b15}, \textbf{KSH}~\cite{b9}. The learning-based hashing methods include \textbf{DPSH}~\cite{b51}, \textbf{DHN}~\cite{b16}, \textbf{CNNH}~\cite{b17}, \textbf{DNNH}~\cite{b33}, \textbf{DSDH}~\cite{b34}. These methods are based on either AlexNet~\cite{b19} or CNN-F~\cite{b59} network architecture. The AlexNet network and CNN-F network have similar network architectures (i.e., They consist of 5 convolutional layers and 2 fully connected layers). As in the traditional hashing methods, we used $\text{DeCAF}_7$ features~\cite{b60}. For learning-based methods, we used raw images as input. In fact, in the past few years, many more advanced networks have been created such as ResNet~\cite{b20}, WRNs~\cite{ba1}. The aim of our paper is to demonstrate a novel technique based on AlexNet that is able to outperform baseline models. If we adopted the advanced networks, we would be unable to know whether the performance gain was given by our \textbf{DAgH} method or by the advanced networks.

We evaluated the image retrieval quality on four metrics:  mean Average Precision (\textbf{mAP}), Precision-Recall curves (\textbf{PR}), Precision curves within Hamming distance 2 (\textbf{P@H}$\mathbf{=}$\textbf{2}), Precision curves with different Number of top returned samples (\textbf{P@N}). For fair comparison, we adopted MAP@1000 for ImageNet and MAP@5000 for other datasets as in~\cite{b34}

\subsection{Implementation Details}
The \textbf{DAgH} method was implemented on Pytorch and batch gradient descent was used to train the network. As shown in Figure~\ref{fig:F2}, our model consists of three networks: an attention network and two hashing networks. We use a very famous attention network, i.e., FCN~\cite{b61} as the base model for the attention network. As discussed in~\cite{b61}, there are three different network models (i.e., FCN-8s, FCN-16s, and FCN-32s). We use the fusing method of FCN-16s to improve performance. Readers can find more details about the attention network in~\cite{b61}. We used AlexNet for the all hashing networks. We fine-tuned convolutional layers and fully-connected layers copied from AlexNet pre-trained on ImageNet and trained the hashing layer $fch$ by back-propagation (BP). As the $fch$ layer is trained from scratch, we set its learning rate to be 10 times that of the lower layers. In our proposed \textbf{DAgH} method, in batch form are used as the input and every two images in the same batch constitute an image pair. The parameters of our proposed \textbf{DAgH} model are learned by minimizing the proposed loss function. The training procedure, i.e., \textbf{DAgH}, is summarized in Algorithm 1.

\textbf{Network Parameters} In our \textbf{DAgH}, the value of hyper-parameter $\nu$ is 50 and $\lambda$ is 0.3. The parameter $\epsilon$ of ATanh follows the empirical value of 0.001 in~\cite{b55}. We use mini-batch Stochastic Gradient Descent (SGD) with 0.9 momentum and the learning rate annealing strategy implemented in Pytorch. The mini-batch size chosen was 32 and the weight decay parameter selected was 0.0005.


\begin{table*}[tp]
\footnotesize
\centering
\begin{threeparttable}
  \centering
  \caption{mean Average Precision (mAP) of Hamming Ranking for Different Number of Bits on the Three Image Datasets.}
  \label{tab:the_overall_performance_mAP}
    \begin{tabular}{l|cccc|cccc|cccc}
    \toprule
    \specialrule{0em}{2pt}{2pt}
    \multirow{2}{*}{Method}&
    \multicolumn{4}{c|}{CIFAR-10}&\multicolumn{4}{c|}{NUS-WIDE}&\multicolumn{4}{c}{ImageNet}\cr
    \cmidrule(lr){2-5}\cmidrule(lr){6-9}\cmidrule(lr){10-13}
    &12 bits&24 bits&32 bits&48 bits&12 bits&24 bits&32 bits&48 bits&12 bits&24 bits&32 bits&48 bits\cr
    \specialrule{0em}{3pt}{3pt}
    \midrule
    \specialrule{0em}{2pt}{2pt}
    SH~\cite{b7}&0.127&0.128&0.126&0.129&0.454&0.406&0.405&0.400&0.185&0.273&0.328&0.395\cr
    \specialrule{0em}{1pt}{1pt}
    ITQ~\cite{b8}&0.162&0.169&0.172&0.175&0.452&0.468&0.472&0.477&0.305&0.363&0.462&0.517\cr
    \specialrule{0em}{1pt}{1pt}
    \hline
    \specialrule{0em}{1pt}{1pt}
    SDH~\cite{b15}&0.285&0.329&0.341&0.356&0.568&0.600&0.608&0.637&0.253&0.371&0.455&0.525\cr
    \specialrule{0em}{1pt}{1pt}
    KSH~\cite{b9}&0.303&0.337&0.346&0.356&0.556&0.572&0.581&0.588&0.136&0.233&0.298&0.342\cr
    \specialrule{0em}{1pt}{1pt}
    \hline
    \specialrule{0em}{1pt}{1pt}
    DHN~\cite{b16}&0.555&0.594&0.603&0.621&0.708&0.735&0.748&0.758&0.269&0.363&0.461&0.530\cr
    \specialrule{0em}{1pt}{1pt}
    CNNH~\cite{b17}&0.429&0.511&0.509&0.522&0.611&0.618&0.625&0.608&0.237&0.364&0.450&0.525\cr
    \specialrule{0em}{1pt}{1pt}
    DNNH~\cite{b33}&0.552&0.566&0.558&0.581&0.674&0.697&0.713&0.715&0.219&0.372&0.461&0.530\cr
    \specialrule{0em}{1pt}{1pt}
    DPSH~\cite{b51}&0.713&0.727&0.744&0.757&0.752&0.790&0.794&0.812&0.143&0.268&0.304&0.407\cr
    \specialrule{0em}{1pt}{1pt}
    DSDH~\cite{b34}&0.726&0.762&0.785&0.803&0.743&0.782&0.799&0.816&0.312&0.353&0.481&0.533\cr
    \specialrule{0em}{1pt}{1pt}
    \hline
    \specialrule{0em}{1pt}{1pt}
    \textbf{DAgH}&\textbf{0.731}&\textbf{0.777}&\textbf{0.809}&\textbf{0.821}&\textbf{0.753}&\textbf{0.791}&\textbf{0.811}&\textbf{0.825}&\textbf{0.322}&\textbf{0.377}&\textbf{0.503}&\textbf{0.551}\cr
    \specialrule{0em}{2pt}{2pt}
    \bottomrule
    \end{tabular}
\end{threeparttable}
\end{table*}

\begin{figure*}
 \centering
  \subfigure[Precision-Recall curve @ 48 bits]{
   \includegraphics[width=2.15in]{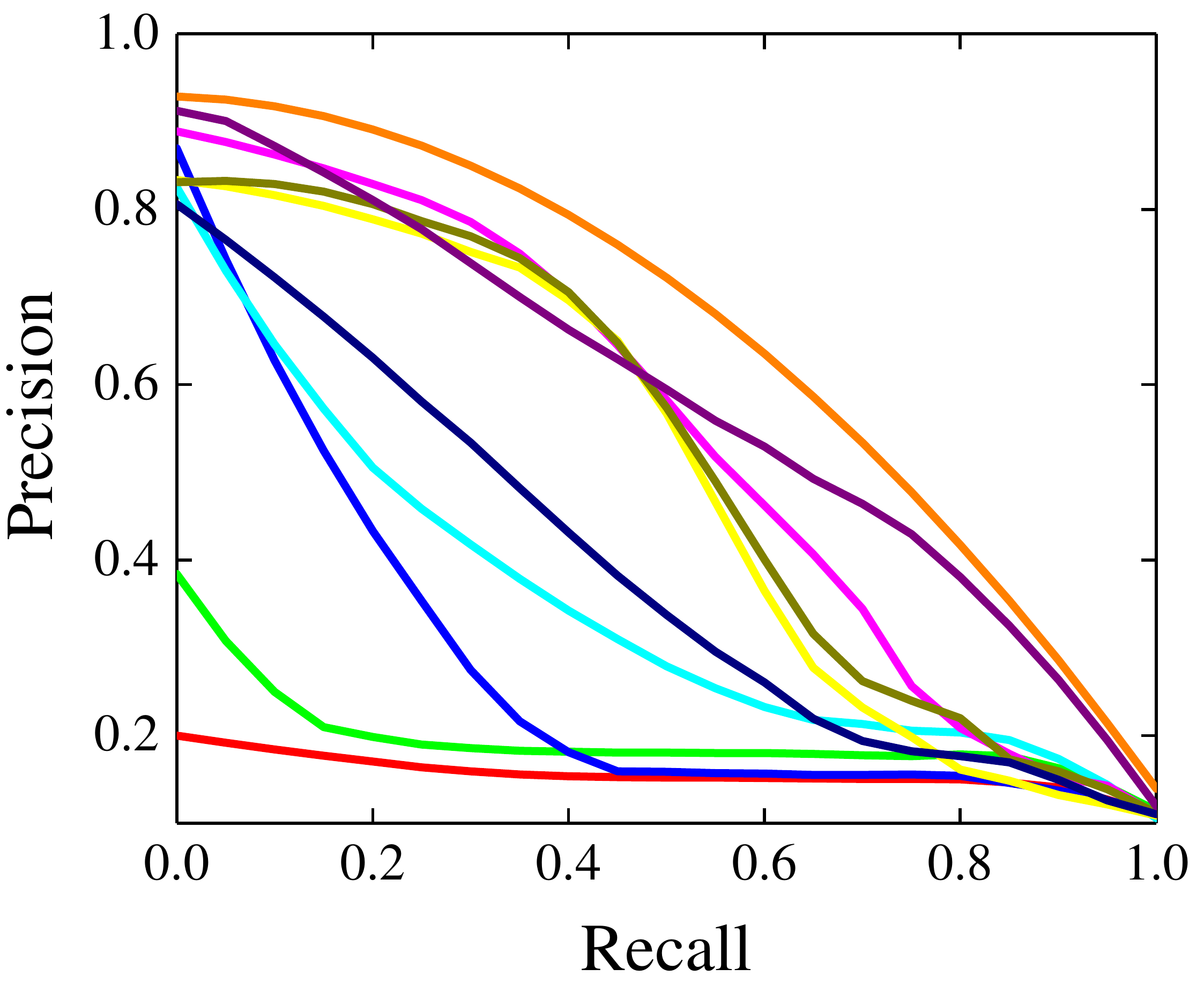}
   }
  \subfigure[Precision within Hamming radius 2]{
 \includegraphics[width=2.13in]{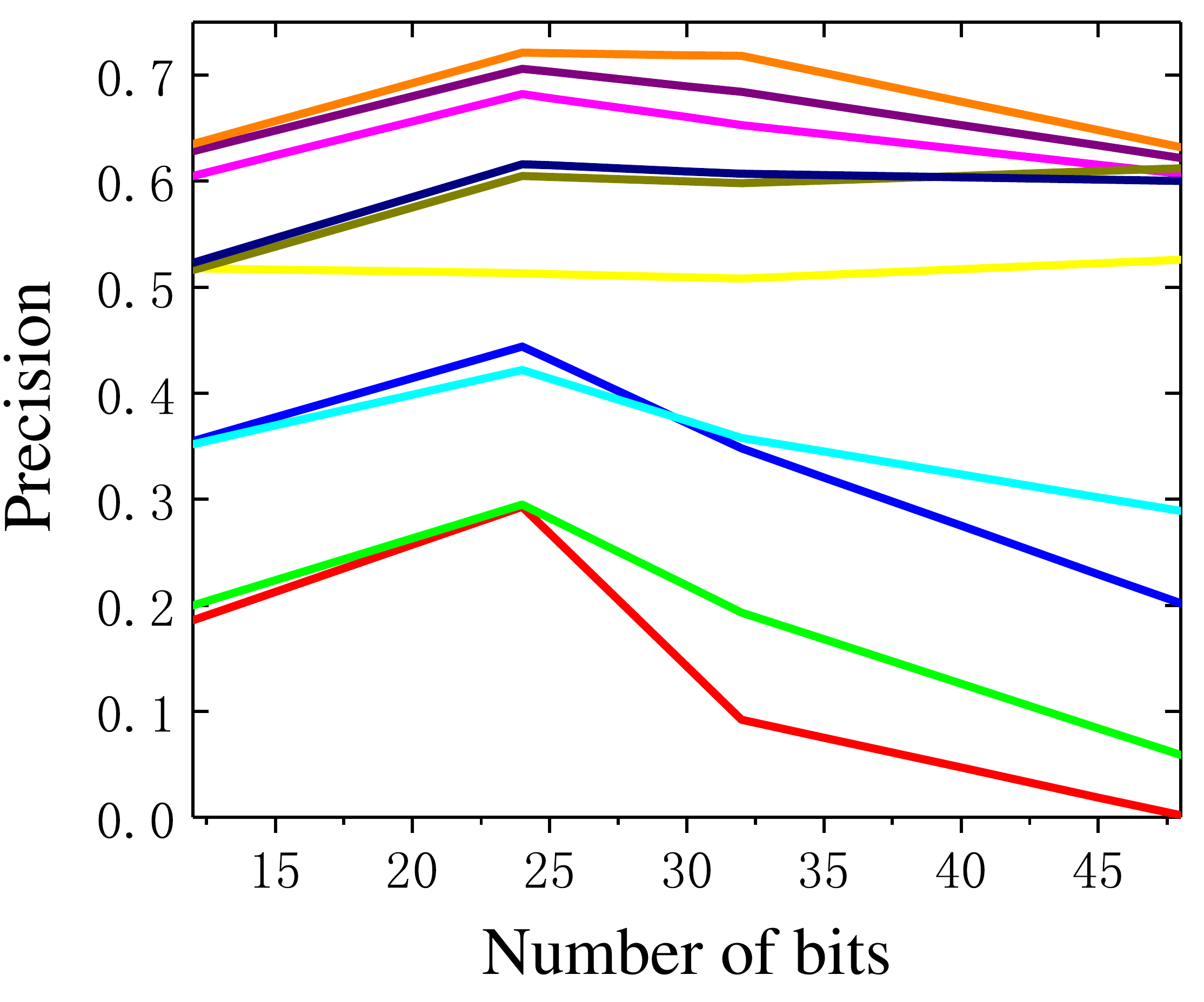}
 }
  \subfigure[Precision curve w.r.t. top-$n$ @ 48 bits]{
   \includegraphics[width=2.42in]{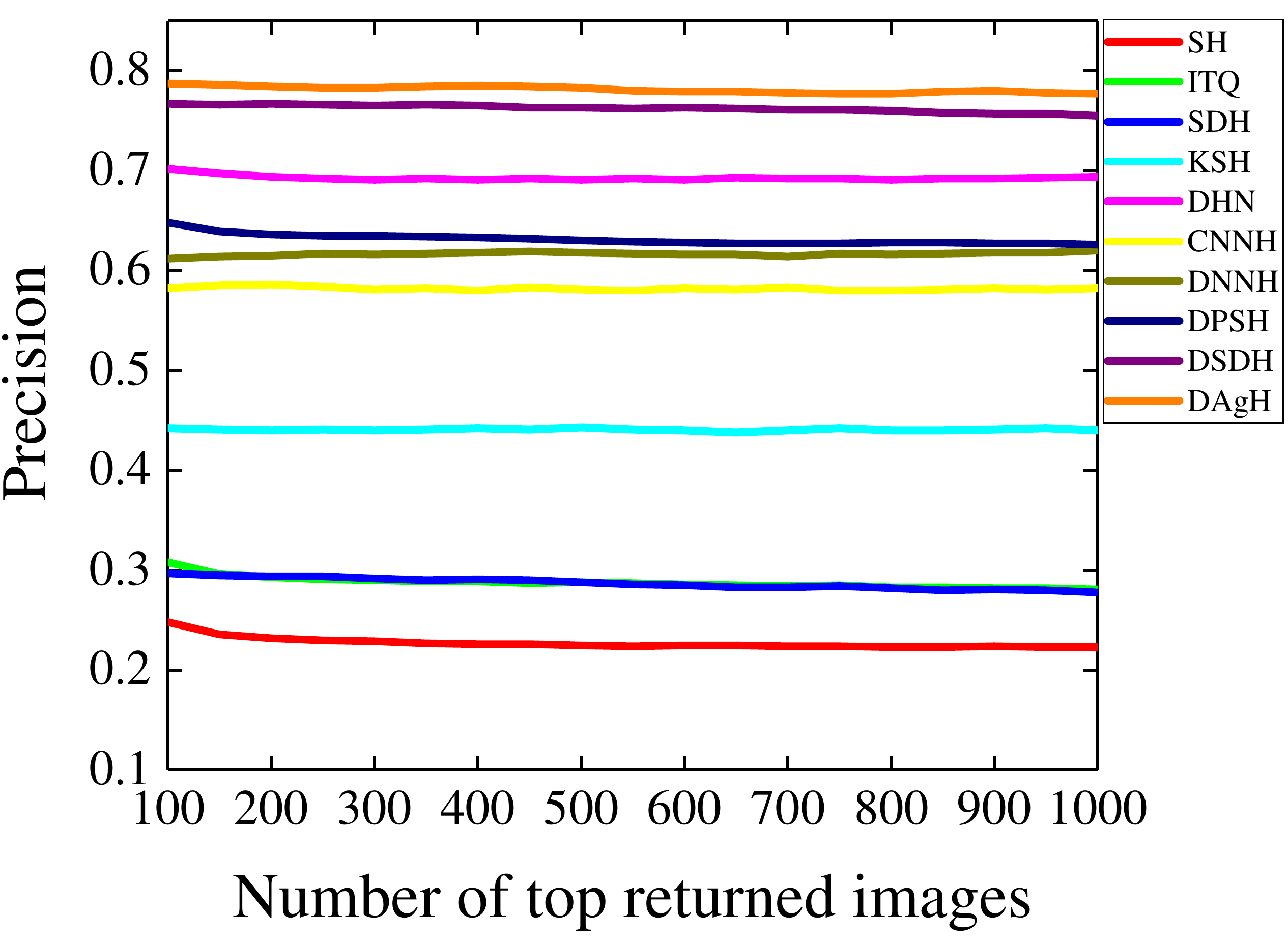}
 }
 \caption{The results of \textbf{DAgH} and comparison methods on the CIFAR-10 dataset under three evaluation metrics.}
 \label{fig:C10} 
 \end{figure*}

\begin{figure*}
 \centering
  \subfigure[Precision-Recall curve @ 48 bits]{
   \includegraphics[width=2.15in]{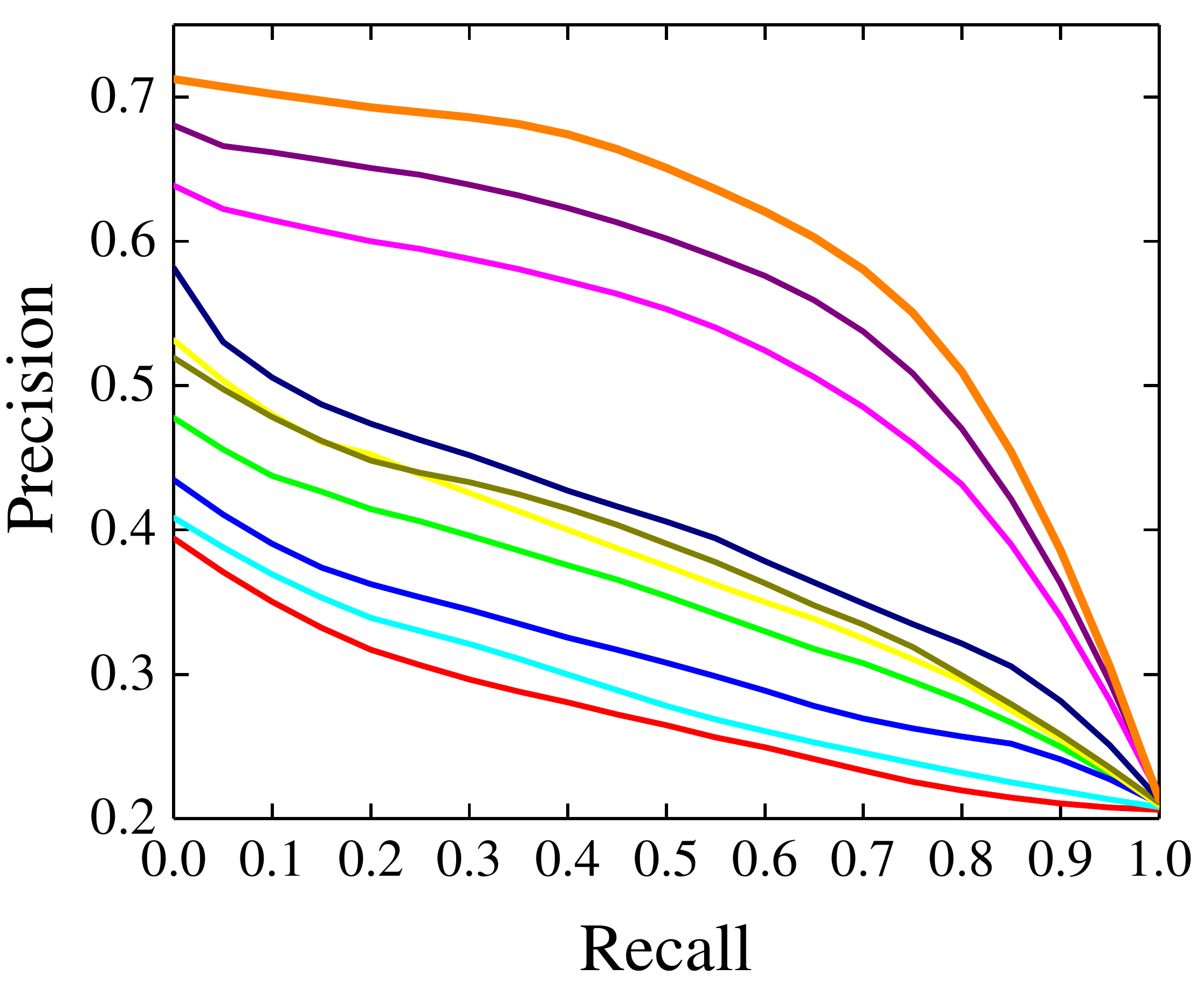}
   }
  \subfigure[Precision within Hamming radius 2]{
 \includegraphics[width=2.12in]{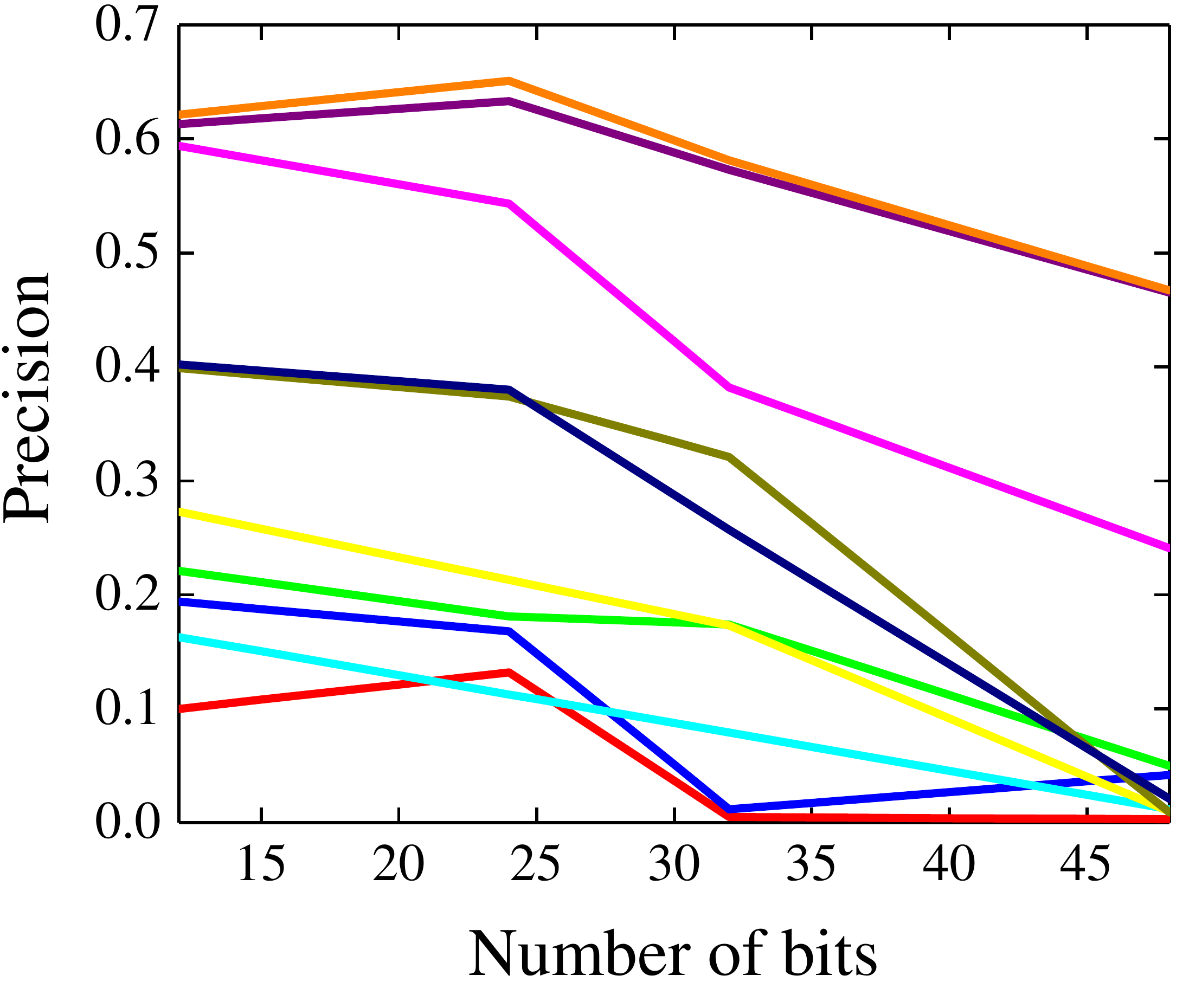}
 }
  \subfigure[Precision curve w.r.t. top-$n$ @ 48 bits]{
   \includegraphics[width=2.43in]{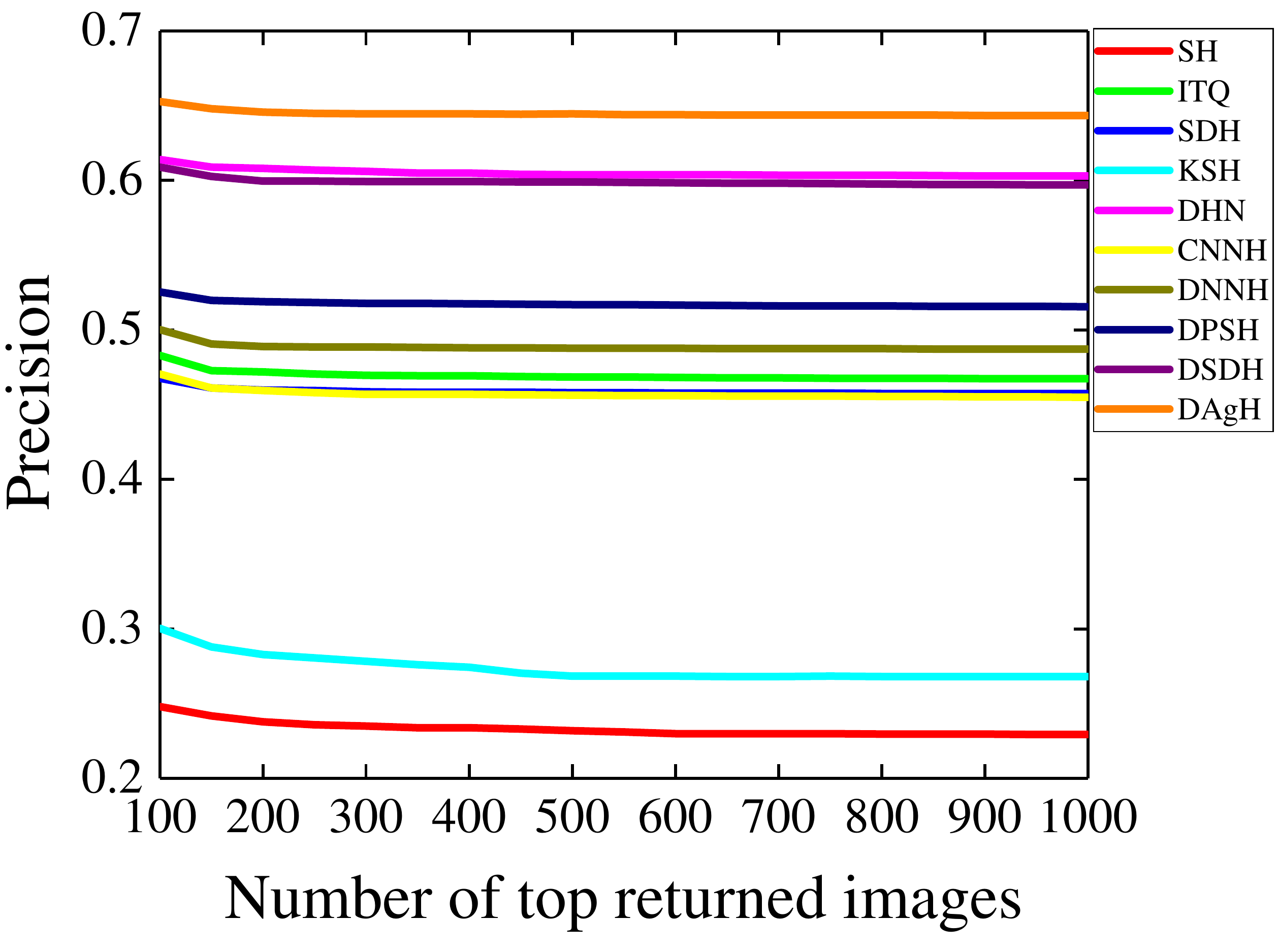}
   }
 \caption{The results of \textbf{DAgH} and comparison methods on the NUS-WIDE dataset under three evaluation metrics.}
 \label{fig:NUS_WIDE} 
 \end{figure*}

\begin{figure*}
 \centering
  \subfigure[Precision-Recall curve @ 48 bits]{
   \includegraphics[width=2.14in]{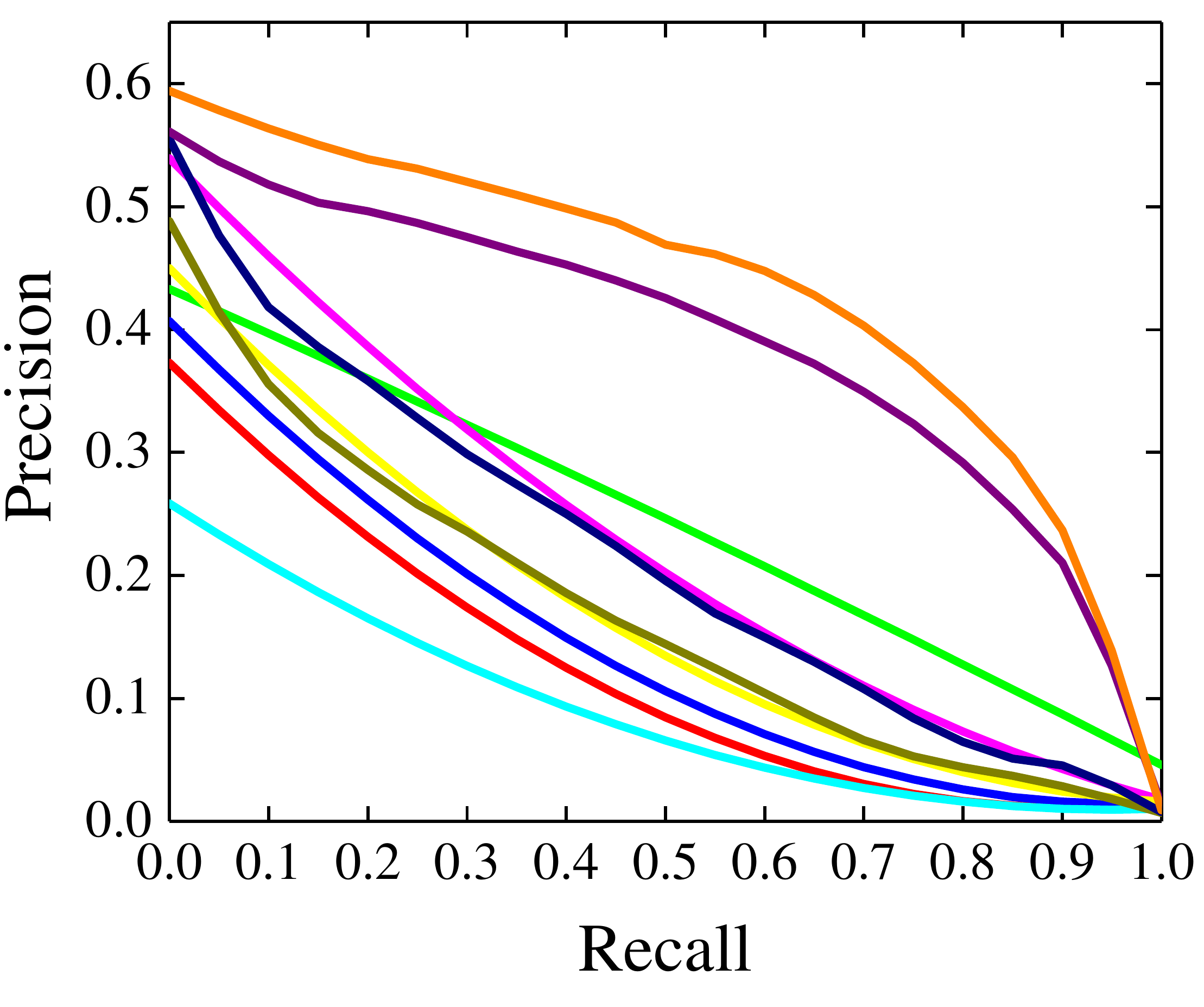}
   }
  \subfigure[Precision within Hamming radius 2]{
 \includegraphics[width=2.13in]{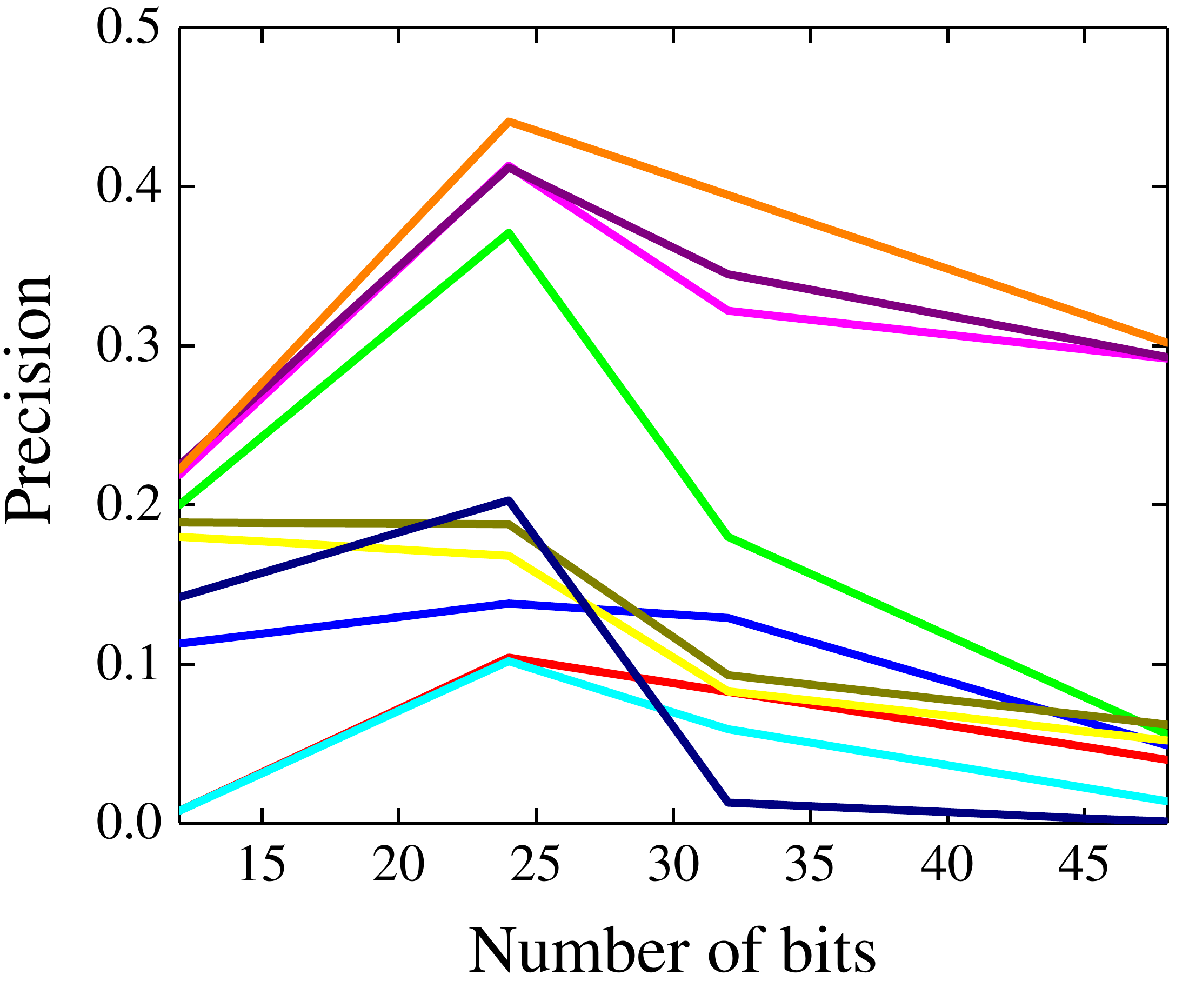}
 }
  \subfigure[Precision curve w.r.t. top-$n$ @ 48 bits]{
   \includegraphics[width=2.44in]{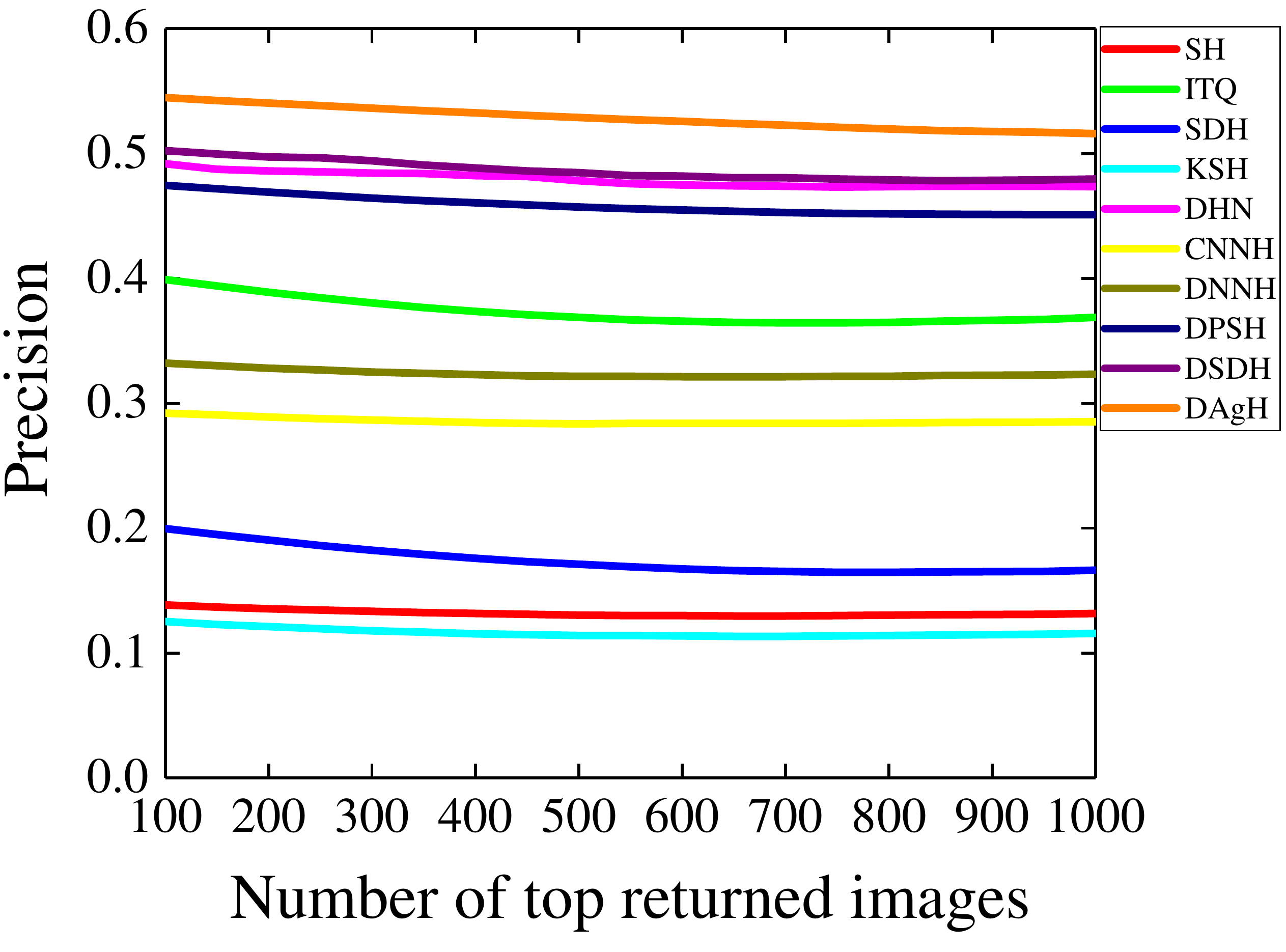}
   }
 \caption{The results of \textbf{DAgH} and comparison methods on the ImageNet dataset under three evaluation metrics.}
 \label{fig:Image} 
 \end{figure*}

\subsection{Results and Discussions}
The mAP results of all methods for different lengths of hash codes on CIFAR-10, NUS-WIDE, and ImageNet are listed in Table~\ref{tab:the_overall_performance_mAP}. Results on CIFAR-10 dataset show that the proposed \textbf{DAgH} method substantially outperforms all other methods against which it was compared. Compared to traditional hashing methods, such as, ITQ, the best shallow hashing method using deep features achieves an absolute boost of 77.83\%, 78.25\%, 78.74\%, and 78.68\% corresponding to different lengths of hash codes, respectively. In addition, most of the learning-based hashing methods perform better than the traditional hashing methods. In particular, DSDH, the state-of-the-art learning-based hashing method, achieves the best performance among all the learning-based methods. Compared to DSDH, our \textbf{DAgH} method can achieve absolute boosts of 0.68\%, 1.90\%, 2.9\%, and 2.2\% in average mAP corresponding to different lengths of hash codes, respectively. Similar to the other hashing methods, we also conducted experiments for large-scale image retrieval. For NUS-WIDE and ImageNet datasets, if two images share at least one same label, they are considered to belong to the same category. The results of experiments using the NUS-WIDE and ImageNet datasets on Table~\ref{tab:the_overall_performance_mAP} show that the proposed \textbf{DAgH} method outperforms the best existing traditional hashing image retrieval methods (i.e., ITQ) by 41.20\% and 5.17\% in average mAP for different lengths of hash codes on these datasets, respectively. Compared to the state-of-the-art learning-based hashing method (i.e., DSDH). We achieve absolute boosts of 1.26\% ,3.64\% in average mAP for different lengths of hash codes on these datasets, respectively. These results demonstrate that our approach can boost the retrieval performance.

We also observe from the Table~\ref{tab:the_overall_performance_mAP} that the gap between the learning-based methods and traditional hashing methods is larger on CIFAR-10 dataset than NUS-WIDE and ImageNet datasets. The reasons are that the number of categories in NUS-WIDE and ImageNet datasets are more than those in CIFAR-10 dataset, and each of the image may contain multiple labels. By carefully comparing the performance of different bits, we found that our proposed method showed a higher degree of performance improvement when tested on at long bits (i.e., 32bits and 48 bits) compared to short bits (i.e, 12bits and 24 bits). This means that our approach can make the hash codes more informative.

An important indicator for evaluating image retrieval performance is Precision within Hamming radius 2 (\textbf{P@H=2}) because such Hamming ranking only require $O(1)$ time for query operations. As shown in Figures~\ref{fig:C10}(b),~\ref{fig:NUS_WIDE}(b), and~\ref{fig:Image}(b), \textbf{DAgH} achieves the highest \textbf{P@H=2} results on all the datasets. In particular, \textbf{P@H=2} of \textbf{DAgH} with 24 bits achieves the best performance. This shows that \textbf{DAgH} can learn more quality hash codes. Norouzi \emph{et al.} ~\cite{b62} show that when generating relatively longer hash codes, the Hamming space will become sparse and few data points will fall within the Hamming ball with a radius of 2. This is why many learning-based hashing methods can achieve good image retrieval performances on short hash codes.

The other important indicators are Precision-Recall curves (\textbf{PR}) and Precision curves with a different Number of top returned samples (\textbf{P@N}). These results are shown in Figures~\ref{fig:C10}(a),~\ref{fig:NUS_WIDE}(a),~\ref{fig:Image}(a) and Figures~\ref{fig:C10}(c),~\ref{fig:NUS_WIDE}(c),~\ref{fig:Image}(c), respectively. We can observe that the performance of our proposed model (\textbf{DAgH}) is better than the models to which it was compared. For example, using the proposed model, more semantic neighbors are retrieved, which is desirable in practical applications. In particular, \textbf{DAgH} achieves stable precision improvement at every recall level test and tests on the number of top images returned, which is very useful for real-world practical systems.

\subsection{Other Analysis}
\textbf{Impact of the first hashing network selection:}
As shown in Figure~\ref{fig:F2}, (the architecture of \textbf{DAgH}), we leverage a hashing network in the first stream framework to generate the attention-guided hash codes from the attention images. Intuitively, the performance of the hashing network could affect the quality of the attention-guided hash codes, i.e., the better the attention-guided hash codes is, the better the performance achieved by the second hashing network. To confirm this, we further design a new variant of \textbf{DAgH}, i.e., \textbf{DAgH-ResNet18}, which adopts ResNet-18 as the first hashing network, instead of AlexNet used in previous experiments. ResNet is a well-known convolutional neural network, and its performance in image processing is better than that of AlexNet. We carried out experiments on the NUS-WIDE dataset. The mAP results are shown in Table~\ref{tab:model_change}. \textbf{DAgH-AlexNet} implies that AlexNet was used in the first stream framework and ResNet-18 was used in \textbf{DAgH-ResNet18}. From table~\ref{tab:model_change}, the following observations were made:
\begin{enumerate}
\item \textbf{DAgH-ResNet18} outperforms \textbf{DAgH-AlexNet} in most cases except in the case of 48 bits. This proves that \textbf{DAgH} can obtain better results by using a first hashing network with better performance.

\item The performance gap between \textbf{DAgH-ResNet18} and \textbf{DAgH-AlexNet} was very small. This indicate that \textbf{DAgH} is not sensitive to attention-guided hash codes, this may be because the information of the attention hash codes is diluted when they guide the generation of new hash codes.
\end{enumerate}

\begin{table}[tp]
\centering
\begin{threeparttable}
  \centering
  \caption{Performance comparison of \textbf{DAgH} with different first hashing networks, i.e, AlexNet and ResNet18.}
  \label{tab:model_change}
    \begin{tabular}{ccccc}
    \toprule
    Method&12 bits&24 bits&32 bits&48 bits\cr
    \midrule
    \specialrule{0em}{2pt}{2pt}
    DAgH-AlexNet&0.753&0.791&0.811&\textbf{0.825}\cr
    \specialrule{0em}{2pt}{2pt}
    DAgH-ResNet18&\textbf{0.769}&\textbf{0.796}&\textbf{0.814}&0.823\cr
    \bottomrule
    \end{tabular}
\end{threeparttable}
\end{table}

\begin{figure}
 \centering
  \subfigure{
  \label{fig:hyper-para:a} 
   \includegraphics[width=1.615in]{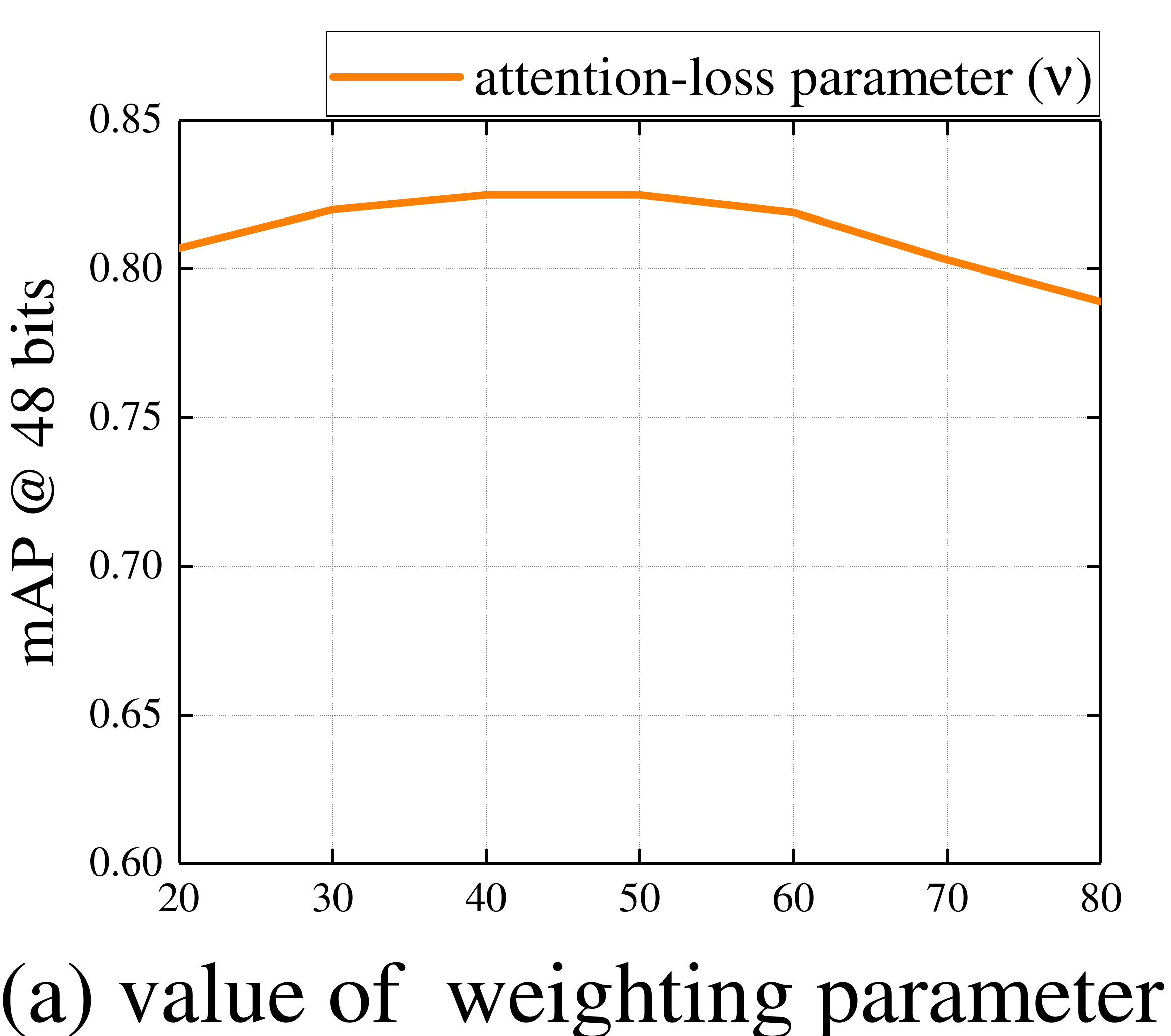}
   }
  \subfigure{
  \label{fig:hyper-para:b} 
 \includegraphics[width=1.572in]{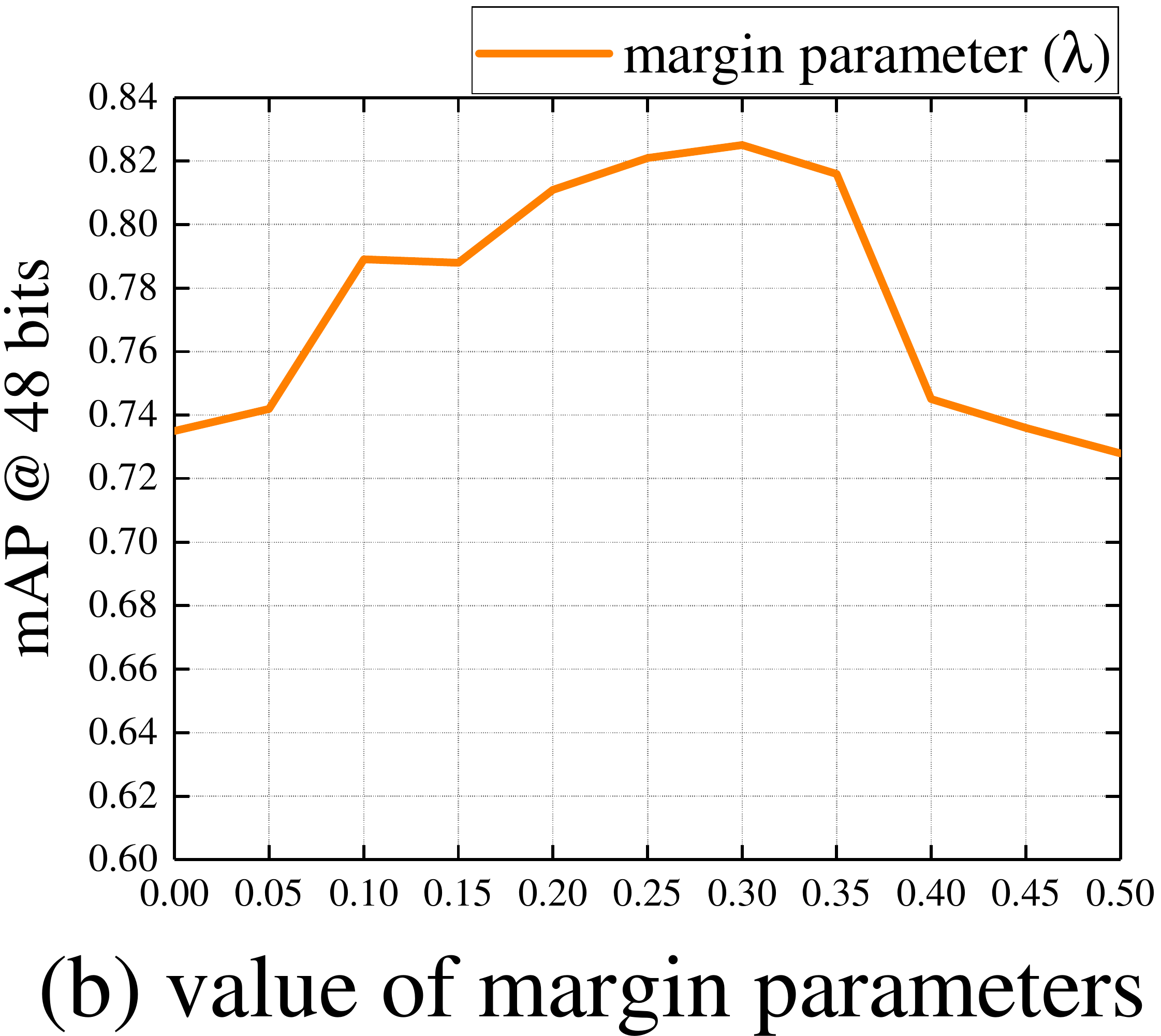}
 }
 \caption{Influence of the hyper-parameters.}
 \label{fig:hyper-para} 
 \end{figure}

\noindent\textbf{Impact of the hyper-parameters:} In this subsection, we analyze the impact of the hyper-parameters, i.e., the value of the attention parameter $\nu$ and the margin parameter $\lambda$. The experiments are conducted on the NUS-WIDE dataset. The value of the attention penalty parameter $\nu$ is selected using values within the range 20 to 80 with a constant step-size of 10 and the margin parameter $\lambda$ is using values within the range 0 to 0.5 with a constant step-size 0.05. Figure~\ref{fig:hyper-para:a} shows that \textbf{DAgH} can achieve good performance on NUS-WIDE dataset within the range $40\leq\nu<60$. As shown in Figure~\ref{fig:hyper-para:b}, the model is sensitive to the value of the margin parameter $\lambda$ and achieved good performance on NUS-WIDE dataset with $0.2\leq\lambda\leq0.35$. This is because according to~\eqref{eq:attention_loss}, if the value of margin is small, the attention loss has a lower impact in punishing the attention network, and as the result, the attention image pair will be similar to the original image pair. If the value of margin is large, the attention loss will affect~\eqref{eq:overall_loss}.

\begin{figure}
    \centering
    \includegraphics[width=8cm]{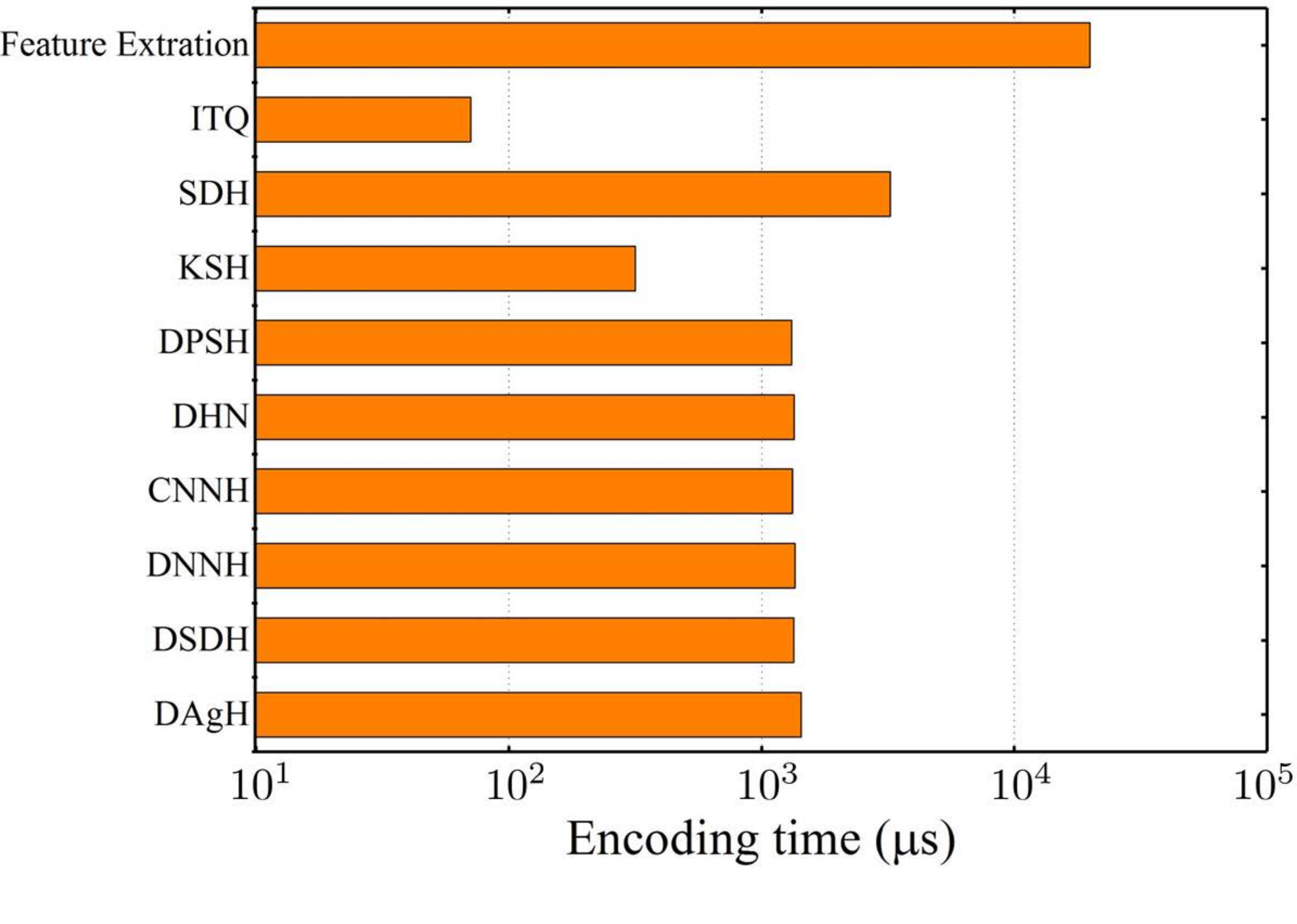}\\
    \caption{The encoding times to encode one new instance (image) of different hashing methods on CIFAR-10 dataset with 48 bits hash codes.}\label{fig:encoding}
  \end{figure}

\noindent\textbf{Encoding Time:} In practical retrieval systems, time efficiency for generating the hash codes for a new instance (image) is an important factor in the evaluation model. In this part, we compare the encoding time of the proposed \textbf{DAgH} method and other baseline hashing methods: \textbf{ITQ}~\cite{b8}, \textbf{SDH}~\cite{b15}, \textbf{KSH}~\cite{b9}, \textbf{DPSH}~\cite{b51}, \textbf{DHN}~\cite{b16}, \textbf{CNNH}~\cite{b17}, \textbf{DNNH}~\cite{b33}, and \textbf{DSDH}~\cite{b34}. Since the input instances are originally raw images, for fair comparison, we take into consideration both the time cost for feature extraction and hashing encoding. We report both the feature extraction time efficiency for traditional hashing methods and the encoding cost for learning-based hashing methods on GPU and the hashing encoding time of traditional hashing methods on CPU. The encoding times (in microseconds, base 10) of involved hashing methods are presented in Figure~\ref{fig:encoding} using a logarithmic scale on the CIFAR-10 dataset with 48 bits hash codes. From Figure~\ref{fig:encoding}, it can be seen that traditional hashing methods such as ITQ, and KSH, actually perform quite decently with encoding times faster than leaning-based hashing methods by an order of magnitude. However, traditional hashing methods require a separate process for feature extraction. When the full process of using a traditional method is put into consideration (feature extraction + traditional hashing method), the encoding time of the traditional methods is much worse than that of leaning-based hashing methods by an order of magnitude. The computing platform is equipped with an Intel 2$\times$ Intel E5-2600 CPU, 128G RAM, and a NVIDIA TITAN Xp 12G GPU. The encoding time basically depends on the adopted neural network model rather than the hashing method. Thus the time varies little with different lengths of hash codes.

\begin{figure}
 \centering
  \subfigure[DAgH]{
  \label{fig:v:a} 
   \includegraphics[width=1.58in]{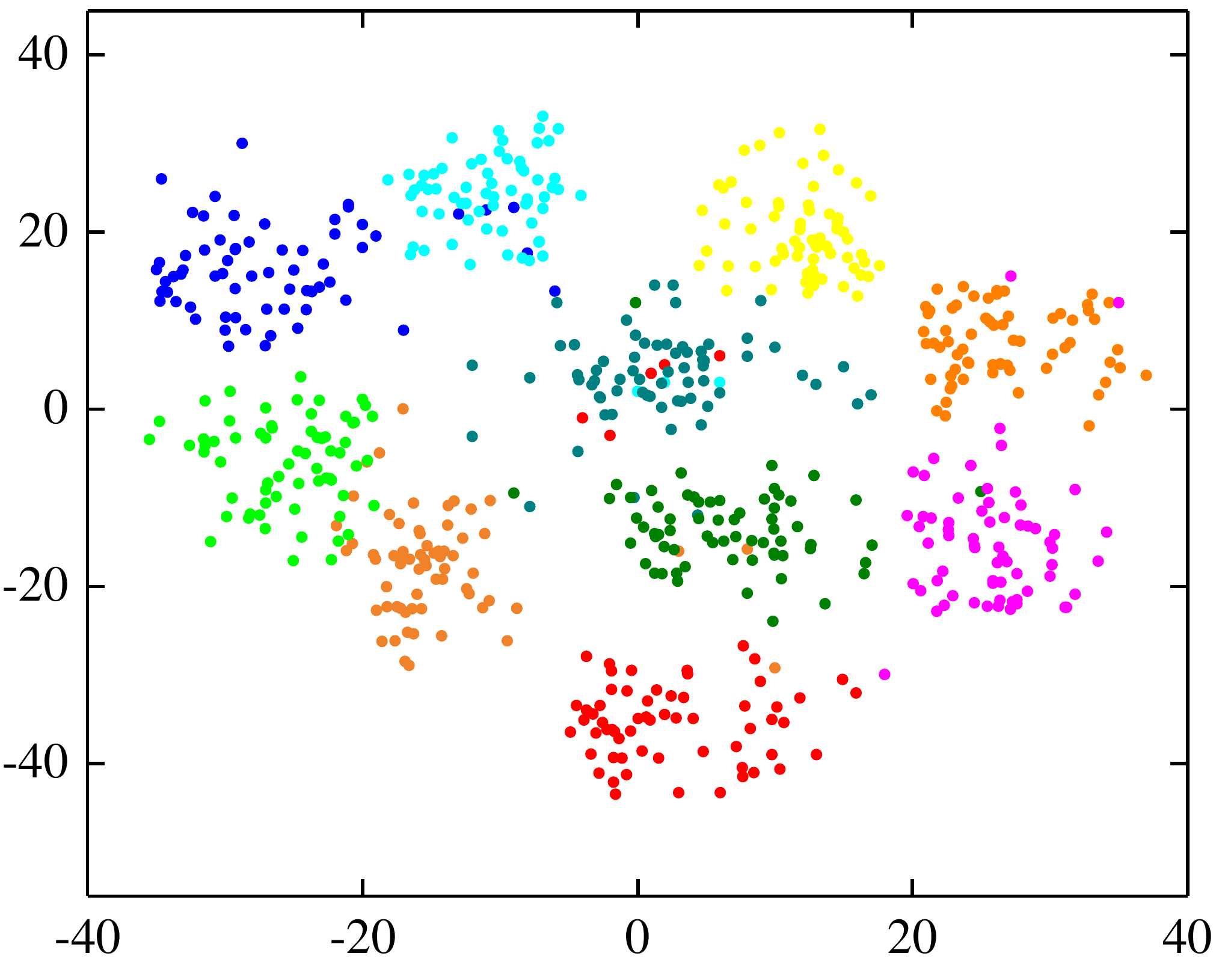}
   }
  \subfigure[DSDH]{
  \label{fig:v:b} 
 \includegraphics[width=1.58in]{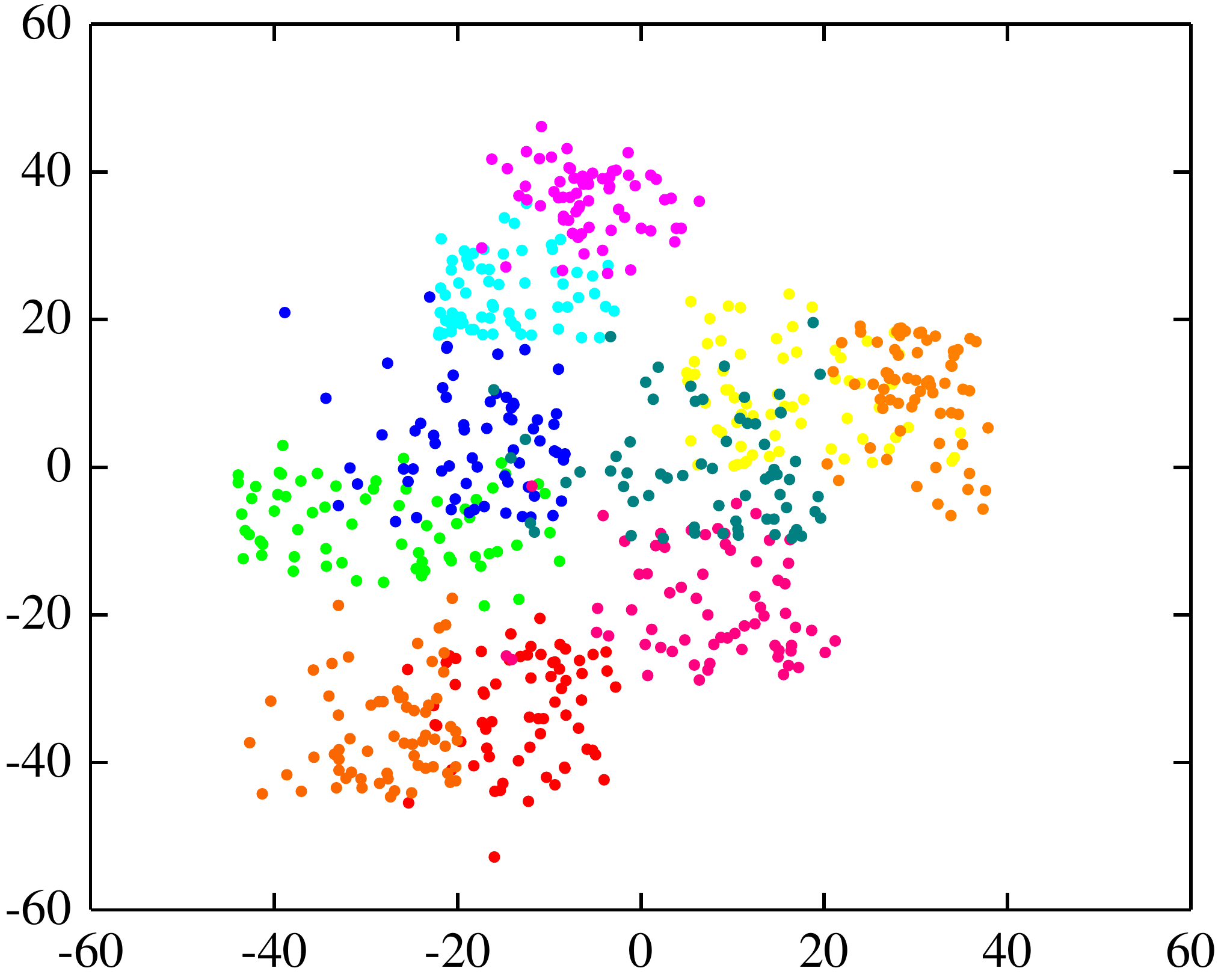}
 }
 \caption{The t-SNE visualization of hash codes learned by \textbf{DAgH} and DSDH.}
 \label{fig:v} 
 \end{figure}

\noindent\textbf{Visualization of Hash Codes:} Figure~\ref{fig:v} shows the t-SNE visualization~\cite{b63} of the hash codes learned by the proposed \textbf{DAgH} method and the best learning-based hashing baseline DSDH on the ImageNet dataset (we sample 10 categories for the case of visualization). We can observe that the hash codes generated by \textbf{DAgH} exhibit clear discriminative structures where the hash codes in different categories are well separated, while the hash codes generated by DSDH do not show such discriminative structures. The results verify that the hash codes learned through the proposed \textbf{DAgH} are more discriminative than those learned by DSDH, enabling more effective image retrieval.

\section{Conclusion}\label{sec:Conclusion_and_Future_Work}
In this paper, we propose a novel attention-guided hashing method for image retrieval, named \textbf{DAgH}. To improve the quality of the generated hash codes, in other words, to address the high correlation problems of the generated hash codes, our method consists of two stream frameworks, which consist of an attention network and two hashing networks. The attention network can automatically mine the key region of an image and generate the attention images. The hashing networks are used to learn semantic-preserving hash codes. The first hashing network generates the attention-guided hash codes from the attention images using pairwise labels to learn the attention-guided hash codes. The second hashing network is then guided by the attention-guided hash codes to generate the final hash codes. On the choice of the hash activation function, the first stream framework uses a continuous ATanh activation function for training and the second stream framework uses a threshold function $sign(\cdot)$. Comprehensive experiments on the three benchmark image retrieval datasets demonstrate that the \textbf{DAgH} outperforms the state-of-the-art methods.

In the future, we plan to extend the self-hashing network to support image retrieval with relative similarity labels, i.e., condense the two stream framework into a single self-training network.


\section*{Acknowledgment}

The authors would also like to thank the associate editor and anonymous reviewers for their comments to improve the paper.


\end{document}